\documentclass{jpp}
\usepackage{subeqn, epsfig}

\ifprodtf \else
  \checkfont{eurm10}
  \iffontfound
    \IfFileExists{upmath.sty}
      {\typeout{^^JFound AMS Euler Roman fonts on the system,
                   using the 'upmath' package.^^J}%
       \usepackage{upmath}}
      {\typeout{^^JFound AMS Euler Roman fonts on the system, but you
                   dont seem to have the}%
       \typeout{'upmath' package installed. JFM.cls can take advantage
                 of these fonts,^^Jif you use 'upmath' package.^^J}%
      }
  \else

  \fi
\fi

\ifprodtf \else
  \checkfont{msam10}
  \iffontfound
    \IfFileExists{amssymb.sty}
      {\typeout{^^JFound AMS Symbol fonts on the system, using the
                'amssymb' package.^^J}%
       \usepackage{amssymb}%
       \let\le=\leqslant

      }{}
  \else
\blacksquare
  \fi
\fi

\ifprodtf \else
  \IfFileExists{amsbsy.sty}
    {\typeout{^^JFound the 'amsbsy' package on the system, using
it.^^J}%
     \usepackage{amsbsy}}
    {\providecommand\boldsymbol[1]{\mbox{\boldmath $##1$}}}
\fi

\newsavebox{\astrutbox}
\sbox{\astrutbox}{\rule[-5pt]{0pt}{20pt}}

\mathchardef\varLambda="0103

\title[Modulated dust-acoustic wave packets in a non-isothermal
background]
  {Modulated dust-acoustic wave packets \\in a plasma with non-isothermal
electrons and ions}

\author[I.\,Kourakis and P.\,K. Shukla]%
  {I.\ns K\ls O\ls U\ls R\ls A\ls K\ls I\ls S$^1$%
   \thanks{On leave from: U.L.B. - Universit\'e Libre de Bruxelles,
Facult\'e des Sciences Apliqu\'ees - C.P. 165/81 Physique
G\'en\'erale, Avenue F. D. Roosevelt 49, B-1050 Brussels, Belgium;
also at: U.L.B. - Universit\'e Libre de Bruxelles, Physique
Statistique et Plasmas C. P. 231, Boulevard du Triomphe, B-1050
Brussels, Belgium; Email: ikouraki@ulb.ac.be.} \ns \and
   \ns P.\ls K.\ns S\ls H\ls U\ls K\ls L\ls A$^2$}

\affiliation{Institut f\"ur Theoretische Physik IV, Fakult\"at
f\"ur Physik und Astronomie Ruhr--Universit\"at Bochum, D--44780
Bochum, Germany
\\[\affilskip]
$^1$ Email: ioannis@tp4.rub.de \  \ \ $^2$ Email: ps@tp4.rub.de}

\pubyear{2004} \volume{XX}
\part{X}
\pagerange{XXX}
\date{26 October 2004; in revised form 10 November 2004}
\begin{document}

\maketitle

\begin{abstract}
Nonlinear self-modulation of the dust acoustic waves is studied,
in the presence of 
non-thermal (non-Maxwellian) ion and electron populations. By
employing a multiple scale technique, a nonlinear
Schr\"odinger-type equation (NLSE) is derived for the wave
amplitude. The influence of non-thermality, in addition to
obliqueness (between the propagation and modulation directions),
on the conditions for modulational instability to occur is discussed. 
Different types of localized solutions (envelope excitations) 
which may possibly occur are discussed,
and the dependence of their characteristics on physical parameters
is traced. The ion deviation from a Maxwellian distribution comes
out to be more important than the electron analogous deviation
alone. Both yield a de-stabilizing effect on (the amplitude of)
DAWs propagating in a dusty plasma with negative dust grains, and thus 
favor the formation of bright- (rather than dark-) type envelope structures 
(solitons) in the plasma. A similar tendency towards amplitude de-stabilization is found 
for the case
of positively charged dust presence in the plasma.
\end{abstract}

\section{Introduction}

Dust contaminated plasmas  have recently received considerable
interest due to their wide occurrence in real charged particle
systems, both in space and laboratory plasmas, and to the novel
physics involved in their description [\cite{Verheest, PSbook}].
An issue of particular interest is the existence of special
acoustic-like oscillatory modes, e.g. the dust-acoustic wave (DAW)
and dust-ion-acoustic wave (DIAW), which were theoretically
predicted about a decade ago [\cite{Rao, SSIAW}] and later
experimentally confirmed [\cite{Barkan, Pieper}]. The DAW, which
we consider herein, is a fundamental electrostatic mode in an
unmagnetized dusty plasma (DP), which has no analogue in an ordinary
\textit{e-i} (electron - ion) plasma. It represents electrostatic
oscillations of mesoscopic size, massive, charged dust grains
against a background of electrons and ions which, given the low
frequency of interest, are practically in a thermalized (i.e.
Boltzmann) equilibrium state. The phase speed of the DAW is much
smaller than the electron and ion thermal speeds, and the DAW
frequency is below the dust plasma frequency.

The amplitude (self-) modulation of the DAse waves has been recently
considered [\cite{AMS} - \cite{IKPKSPS}], by means of the
reductive perturbation formalism [\cite{redpert}]. A study of the
modulational stability profile of DA waves has shown that
long/short wavelength DA waves are stable/unstable
against external perturbations. The respective parameter regions
are associated with the occurrence of 
localized envelope excitations of the dark/bright
type (i.e. voids/pulses). Obliqueness in perturbations are found shown to
modify this dynamic behaviour [\cite{AMS} - \cite{IKPKSPS}].

Allowing for a departure from Boltzmann's distribution for the
electrostatic background ({\emph{non-thermality}}) has been shown
to bear a considerable effect on electrostatic plasma modes.
Inspired by earlier works on the ion-acoustic (IA) solitary waves
[\cite{Cairns}], recent studies have shown that the presence of a
non-thermal ion velocity distribution may modify the nonlinear
behaviour of the DA waves, affecting both the form and the conditions
for the occurrence of the DA solitons [\cite{Gill, Singh}]. Also, the
self-modulation of IA waves was recently shown to be affected by the
electron non-thermality [\cite{Jukui-Tang}]. However, no study has
been carried out of the effect of a non-thermal ion and/or
electron background population on the modulation properties of the DA
waves. This paper aims at filling this gap.

\section{The model}

Let us consider the propagation of the dust-acoustic waves in an
unmagnetized dusty plasma. The mass and charge of dust grains (both
assumed constant, for simplicity) will be denoted by $m_d$ and $q_d
= s \, Z_d e$, where $s = {\rm{sgn}}q_d \equiv q_d/|q_d|$ denotes
the sign of the dust charge ($= \pm 1$). Similarly, we have: $m_i$
and $q_i = + Z_i e$ for ions, and $m_e$ and $q_e =  - e$ for
electrons. The typical DAW frequency will be assumed much lower
than the (inverse of the) characteristic dust grain charging time
scale, so charge variation effects (known to lead to collisionless
wave damping) will be omitted here.

The basis of our study includes the moment - Poisson system of
equations for the dust particles and non-Boltzmann distributed
electrons and ions. The dust number density $n_d$ is governed by
the continuity equation
\begin{equation}
\frac{\partial n_d}{dt} + \nabla \cdot (n_d \,\mathbf{u}_d)= 0 \,
, \label{densityequation}
\end{equation}
and the dust mean velocity $\mathbf{u}_d$ obeys
\begin{equation}
\frac{\partial \mathbf{u}_d}{dt} + \mathbf{u}_d \cdot \nabla
\mathbf{u}_d \, = \, - \frac{q_d}{m_d}\,\nabla \,\Phi
\end{equation}
where $\Phi$ is the electric potential. The dust pressure dynamics
(i.e. the dust temperature effect) is omitted within the cold dust fluid
description. The system is closed by
Poisson's equation
\begin{equation}
\nabla^2 \Phi \, =\, - 4 \pi\, e \,(n_i \, Z_i - n_e + s\, n_d \,
Z_d) \, . \label{Poisson1}
\end{equation}
Overall neutrality, viz. \[ n_{i, 0} \, Z_i - n_{e, 0} + s\, n_{d,
0} \, Z_d = 0
\] is assumed at equilibrium.

\subsection{Non-thermal background distribution(s)}

Following the model of Cairns \textit{et al.} [\cite{Cairns}], the
non-thermal velocity distribution for $e^-$ (electrons) and $i^+$
(ions) is taken to be
\begin{equation}
f_{s'}(v; a_{s'}) = \frac{n_{s', 0}}{\sqrt{2 \pi v_{th, s'}^2}}
\frac{1 + a_{s'} v^4/v_{th, s'}^4 }{1 + 3 a_{s'}} \, \exp(- v^2/2
v_{th, s'}^2) \, ,\label{Cairns-model1}
\end{equation}
where $n_{s', 0}$, $T_{s'}$ and $v_{th, s'} = (k_B
T_{s'}/m_{s'})^{1/2}$ denote the equilibrium density, temperature
and thermal speed of the species $s' \in \{1, 2 \} \equiv  \{
i, e\}$, respectively. The real parameter $a_{s'}$ expresses the
deviation from the Maxwellian state (which is recovered for
$a_{s'} = 0$). Integrating over velocities, Eq.
(\ref{Cairns-model1}) leads to
\begin{equation}
n_{s'} = n_{s', 0} (1 + \sigma_{s'} b_{s'} \Phi' + b_{s'}
{\Phi'}^2)\, \exp({- \sigma_{s'} \Phi'}) \label{Cairns-model2}
\end{equation}
[\cite{Cairns}], where $\sigma_{1/2}= \sigma_{i/e} = +1/-1$ \ and
$b_{s'} = 4 a_{s'}/(1 + 3 a_{s'})$; \ the normalized potential
variables are $\Phi'_{s'} = Z_{s'} e \Phi/k_B T_{s'}$,\ where
$Z_{1/2}=Z_{i}/1$ is the ion/electron charge state.

\subsection{Reduced model equations - weakly nonlinear oscillations}

The above system of evolution equations for the dust fluid can be
cast into the reduced (dimensionless) form:
\begin{equation}
\frac{\partial n}{\partial t} + \nabla (n \,\mathbf{u}) =  0
\label{eq1} \, ,
\end{equation}
and 
\begin{equation}
\frac{\partial \mathbf{u}}{\partial t} + \mathbf{u} \cdot \nabla
\mathbf{u} \, = \, - s\,\nabla \phi \, \label{eq2}
\end{equation}
where the particle density $n_d$, mean fluid velocity $u_d$,
and electric potential $\Phi$ are scaled as: $n = n_d/n_{d, 0}$,
$u = u_d/c_d$,
and $\phi = |q_d| \Phi/(k_B T_{eff})$, where $n_{d, 0}$ is the
equilibrium dust density; the effective temperature $T_{eff}$ is
related to the characteristic dust speed $c_d \equiv (k_B
T_{eff}/m_d)^{1/2} = \omega_{p, d} \lambda_{D, eff}$, defined
below ($k_B$ is Boltzmann's constant). Time and space variables are scaled
over the characteristic dust period (inverse dust plasma frequency)
$\omega_{p, d}^{-1} = (4 \pi n_{d, 0} q_{d}^2/m_d)^{- 1/2}$, and
the effective Debye length, defined as $\lambda_{D, eff} =
[(1-b_e) \lambda_{D, e}^{-2}+ (1-b_i) \lambda_{D, i}^{-2}]^{-1/2}
$ [where $\lambda_{D, e/i} = (k_B T_{e/i}/4 \pi n_{{e/i}, 0}
q_{{e/i}}^2)^{1/2}$ is the Debye length for species $e/i$; the
non-thermality parameters $b_{e/i}$ are defined below].
The (unperturbed) dust Debye length $\lambda_{D, d} =
(\lambda_{D, e}^{-2}+ \lambda_{D, i}^{-2})^{-1/2}
$ is also defined. 

Near equilibrium ($\phi \ll 1$), Poisson's Eq. (\ref{Poisson1}) becomes:
\begin{equation}
\nabla^2 \phi \,  \approx \, \phi - \alpha \,\phi^2 + \alpha'
\,\phi^3 - s \,(n - 1)\, . \label{Poisson2}
\end{equation}
Note that the right-hand side in Eq. (\ref{Poisson2}) cancels at
 equilibrium.
Here,
\[ \alpha = \biggl( \frac{Z_i}{\lambda_{D, i}^2 T_i} -
\frac{1}{\lambda_{D, e}^2 T_e} \biggr) \,\frac{\lambda_{D, eff}^2
T_{eff}}{2 Z_d} \]
and
\[ \alpha' = \biggl[ \frac{(1 + 3 b_i)\,
Z_i^2}{\lambda_{D, i}^2 T_i^2} + \frac{1 + 3 b_e}{\lambda_{D, e}^2
T_e^2} \biggr] \, \frac{\lambda_{D, eff}^2 T_{eff}^2}{6 Z_d^2} \,
.
\] For $T_e \gg T_i$, one may retain the approximate expressions:
$\alpha \approx - s (1 - \mu)/[2 (1 - b_i^2)^2]$ and $\alpha'
\approx (1+3 b_i) (1 - \mu)^2/[6 (1 - b_i^2)^3]$; also,
$\lambda_{D, eff}\approx \lambda_{D, i} (1 - b_i)^{-1/2}$ in this
case (notice that the dependence on the electron parameters
disappears in this approximation). We have defined the dust
parameter $\mu = n_{e, 0}/(Z_i n_{i, 0}) = 1 + s Z_d n_{d, 0}/(Z_i
n_{i, 0})$; see that $\mu$ is lower (higher) than unity for
negative (positive) dust charge, i.e. for $s = -1$ ($+1$). 

The dust parameter $\mu$, together with the temperature ratio 
$t_i \equiv T_i/T_e$, are the (dimensionless) physical parameters essentially affecting our problem. For instance, see that $\lambda_{D, d}$ (defined above) may be expressed as: $\lambda_{D, d} = \lambda_{D, e} \,
\{(t_i \mu/Z_i)/[1+(t_i \mu/Z_i)]\}^{1/2}$.  

The system of the evolution equations (\ref{eq1}) -
(\ref{Poisson2}) will be the basis of the analysis that follows.

\section{Perturbative analysis}

 Following the reductive perturbation
technique [\cite{redpert}], we define the state vector $\mathbf{S}
= \{n, \mathbf{u}, \phi\}$, whose evolution is governed by Eqs.
(\ref{eq1}) - (\ref{Poisson2}), and then expand all of its
elements in the vicinity of the equilibrium state
${\mathbf{S}}^{(0)} = (1, \mathbf{0}, 0)^T$,
 viz.
 ${\mathbf{S}} = {\mathbf{S}}^{(0)} + \epsilon {\mathbf{S}}^{(1)} +
 \epsilon^2 {\mathbf{S}}^{(2)} + ...$,
 where $\epsilon \ll 1$ is a
 smallness parameter.
We assume that $ S{j}^{(n)} \,= \, \sum_{l=-\infty}^\infty \,S_{j,
l}^{(n)}(X, \, T) \, e^{i l (\mathbf{k r} - \omega t)}$ (for $j=1,
2, ...$; we impose $S_{j,-l}^{(n)} = {S_{j, l}^{(n)}}^*$, for
reality), thus allowing the wave amplitude to depend on the
stretched (\emph{slow}) coordinates $ X \,= \, \epsilon (x -
\lambda \,t)$ and $T \,= \, \epsilon^2 \, t$, where $\lambda$
($\in \Re$) (bearing dimensions of velocity) will be determined
later. Note that the choice of direction of the propagation
remains arbitrary, yet modulation is allowed to take place in an
oblique direction, characterized by a pitch angle $\theta$. Having
assumed the modulation direction to define the $x-$ axis, the wave
vector $\mathbf{k}$ is taken to be \( \mathbf{k} = (k_x, \,
k_y) = (k\, \cos\theta, \, k\, \sin\theta) \). 

According to the above considerations, the derivative operators in
the above equations are treated as 
\[
\frac{\partial}{\partial t} \rightarrow \frac{\partial}{\partial
t} - \epsilon \, \lambda \, \frac{\partial}{\partial \zeta} +
\epsilon^2 \, \frac{\partial}{\partial \tau} \, ,
\]
\[
\nabla \rightarrow \nabla + \epsilon \, \hat x \,
\frac{\partial}{\partial \zeta} \, ,
\]
and
\[
\nabla^2 \rightarrow \nabla^2 + 2 \epsilon \,
\frac{\partial^2}{\partial x \partial\zeta} + \epsilon^2 \,
\frac{\partial^2}{\partial\zeta^2} \, ,
\]
i.e. explicitly
\[
\frac{\partial }{\partial t} \, A_l^{(n)} \, e^{i l \theta_1} =
\biggr( - i l \omega \, A_l^{(n)} \, - \epsilon \, \lambda \,
\frac{\partial A_l^{(n)}}{\partial \zeta} + \epsilon^2 \,
\frac{\partial A_l^{(n)} }{\partial \tau} \biggr) \, e^{i l
\theta_1} \, ,
\]
\[
\nabla \, A_l^{(n)} \, e^{i l \theta_1} = \biggr( i l \mathbf{k}
\, A_l^{(n)} \, + \epsilon \, \hat x \, \frac{\partial
A_l^{(n)}}{\partial \zeta}  \biggr) \, e^{i l \theta_1}  \, ,
\]
and
\[
\nabla^2 A_l^{(n)}\, e^{i l \theta_1} = \biggr( - l^2 k^2 \,
A_l^{(n)} \, + 2 \epsilon \, i l k_x \, \frac{\partial
A_l^{(n)}}{\partial\zeta} + \epsilon^2 \, \frac{\partial^2
A_l^{(n)}}{\partial\zeta^2} \biggr) \, e^{i l \theta_1}
\]
for any $A_l^{(n)}$ of the components of $S_l^{(n)}$.

\subsection{First harmonic amplitudes}

Iterating the standard perturbation scheme [\cite{redpert}], we
obtain the first harmonic amplitudes (to order $\sim \epsilon^1$)
\begin{eqnarray}
n_1^{(1)} \, = \, s ({1 + k^2})\,  \psi \, , \qquad u_{1, x}^{(1)}
\, = \frac{\omega}{k} \cos\theta \, n_1^{(1)} \, ,\qquad u_{1,
y}^{(1)} \, = \frac{\omega}{k} \sin\theta\, n_1^{(1)} \, ,
\label{1st-order-corrections}
\end{eqnarray}
which may be expressed in terms of a sole state variable, e.g. the
potential correction $\phi_1^{(1)} \equiv \psi$. The well-known DA
wave (reduced) dispersion relation \( \omega^2\, = {k^2}/({k^2 +
1})\) is also obtained, as a compatibility condition.

The amplitudes of the 2nd and 0th (constant) harmonic corrections
are obtained in order $\sim \epsilon^2$ (the lengthy expressions
are omitted for brevity). Furthermore, the condition for
suppression of the secular terms leads to the compatibility
condition: \( \lambda \,   = {\partial \omega}/{\partial k_x} =
\omega'(k) \cos\theta =  k \cos\theta/ [\omega (1 + k^2)^2] \);
therefore, $\lambda$ bears the physical meaning of a group
velocity \emph{in the direction of amplitude modulation} ($\sim
\hat x$).

\subsection{The envelope evolution equation}

The potential correction $\psi$ obeys an explicit compatibility
condition in the form of the \emph{nonlinear Schr\"odinger--type
equation} (NLSE)
\begin{equation}
i\, \frac{\partial \psi}{\partial T} + P\, \frac{\partial^2
\psi}{\partial X^2} + Q \, |\psi|^2\,\psi = 0 \, .  \label{NLSE}
\end{equation}
The {\em dispersion coefficient} $P$ is related to the curvature
of the dispersion curve as \( P \,  = \, {\partial^2 \omega}/{2
\partial k_x^2} \,= \, \bigl[ \omega''(k) \,
\cos^2\theta \, + \omega'(k) \, {\sin^2\theta}/{k} \bigr]/2 \);
the exact form of P reads
\begin{equation}
P(k) \,  =\, \frac{1}{2\,\omega} \,
\biggl(\frac{\omega}{k}\biggr)^4\, \bigl[ 1 - (1 + 3 \,\omega^2)\,
\cos^2\theta \bigr] \, . \label{Pcoeff}
\end{equation}
The {\em nonlinearity coefficient} $Q$, due to carrier wave
self-interaction, is given by a complex function of $k$, $\alpha$
and $\alpha'$. Distinguishing different contributions, $Q$ can be
split into three distinct parts, viz.
\begin{equation}
Q = \, Q_0 \, +\, Q_1 \, +\, Q_2  \, , \label{Qstructure}
\end{equation}
where
\begin{eqnarray}
Q_0 &=& \, + \, \frac{1}{2 \omega}\, \frac{1}{(1 + k^2)^2}\,
\frac{1}{1 - \lambda^2}\,\nonumber \\ && \times \biggl\{ k^2 \,
                        \biggl[
    \,
        \bigl[ 3 + 6 k^2 + 4 k^4 + k^6
                        + 2 \,\alpha \,\bigl(s \,(2 k^2 + 3) + \,2
\,\alpha \,\lambda^2
                \bigr) \bigr]
\nonumber \\
&&
      \qquad \qquad \qquad  + \,
                     \beta \, (2 + 4 k^2 + 3 k^4 + k^6 + 2 s \alpha
\beta ) \, \cos 2\theta \biggr] \nonumber \\ &&
      + \, 2 \, (1 + k^2)^4 \,\omega^2 \, \cos^2\theta
\nonumber \\ &&
       + \, k \, (1 + k^2) \, \biggl[k^2 + \omega^2 \, (1 + k^2)
\biggr]\,
       \frac{\lambda}{\omega} \,
(1+ k^2 + 2 s \alpha \beta)  \,
       \cos\theta \biggr\}
\, , \label{Q0coeff}
\\ Q_1 &=&
\, \frac{3 \, \alpha'}{2 \,\omega}\, \frac{k^2}{(1 + k^2)^2} \,
,\label{Q1coeff}
\\
Q_2 &= & \, - \, \frac{1}{12}\,\frac{1}{\omega}\,
 \frac{1}{k^2 \, (1 + k^2)^2} \,
\nonumber \\ &&
 \times \biggl\{  2 \, k^2 \,
                  \bigl[ 5 \,s \,\alpha \,  (1 + k^2)^2 +
                  \, 2 \alpha^2
              + \, 3 (1 +
k^2)^3 \, (1 + 3 k^2) \bigr] \nonumber \\ && \qquad \qquad \qquad
\qquad + \, (1 + k^2)^3 \, \omega^2 \, (3 + 9 k^2 + 6 k^4 + 2 s
\alpha) \, \biggr\} \, .
 \label{Q2coeff}
\end{eqnarray}
One may readily check, yet after a tedious calculation, that
expressions (\ref{Q0coeff}) and (\ref{Q2coeff}) reduce to (53) and
(54) in Ref. \cite{AMS} (for a Maxwellian distribution, i.e. setting
$a_1=a_2=0$ in all formulae above); notice however, that the term
$Q_1$ was absent therein. Note that only $Q_0$ (which is related
to self-interactions due to the zeroth harmonic) depends on the
angle $\theta$ (and is, in fact, an even, $\pi-$periodic function
of $\theta$).

\section{Stability profile -- envelope excitations}

The evolution of a wave whose amplitude obeys Eq. (\ref{NLSE})
essentially depends on the sign of the product $P Q$
(see, e.g., in Ref. \cite{Hasegawa} for details; also see in Refs.
\cite{IKPKSPS} and \cite{SPIG}) which may be numerically
investigated in terms of the physical parameters involved.

For $P Q > 0$, the DA wave is modulationally \emph{unstable} and
may either \emph{collapse}, when subject to external
perturbations, or evolve into a series of \emph{bright}-type
localized envelope wavepackets, which represent a
\emph{pulse}-shaped solution of the NLSE (\ref{NLSE}). This type
of solution is depicted in Fig. \ref{figure2}.

For $P Q < 0$, on the other hand, the DA wave is \emph{stable} and
may propagate as a \emph{dark/grey}-type envelope wavepacket, i.e.
a propagating localized envelope \emph{hole} (a \emph{void})
amidst a uniform wave energy region. Notice the qualitative
difference between the dark and grey solutions (depicted in Fig.
\ref{figure3}a and b, respectively): the potential vanishes in the
former, while it remains finite in the latter.

The exact analytical expressions for different types of
envelope excitations (depicted in Figs. \ref{figure2} and
\ref{figure3}) are summarized in Refs. \cite{IKPKSPS} and
\cite{SPIG}; more details can be found in Ref. \cite{Fedele}.
 We note that, in either case (i.e. for positive or negative  product
$P Q$),
 the localized excitation \emph{width}
$L$ and \emph{maximum amplitude} $\psi_0$ satisfy $L \psi_0 = (2
P/Q)^{1/2} = {\rm{constant}}$. The dynamics, therefore, essentially
depends on the quantity $\eta = P/Q$, whose sign (magnitude) will
determine the wave's stability profile and the type
(characteristics, respectively) of the localized envelope
excitations which may occur.

\section{Numerical analysis}

According to the above analysis, both the DAW stability profile
and the type of DA envelope excitations possibly occurring in the
plasma are determined by the \emph{sign} of the product of the
NLSE coefficients $P$ and $Q$, which is essentially a function of
the wavenumber $k$ (normalized by $k_D \equiv \lambda_{D, d}^{-1}$ 
[\cite{comment}]), the dust parameter $\mu$, the
temperature ratio $T_i/T_e \equiv t_i$ and the non-thermality
parameters $a_{i, e}$, in addition
to the modulation angle $\theta$. The exact expressions obtained
for $P$ and $Q$ may now be used for the numerical investigation of
the wave's modulational stability profile.

In the figures provided here (see Figs. \ref{figure4} - \ref{figure9}), the
black/white regions represent a negative/positive $P Q$ sign,
implying DAW being modulationally stable/ unstable (and possibly
propagating in the form of dark/bright type excitations,
respectively).

Throughout this Section, we have used the set of representative
values: $t_i = T_i/T_e = 0.1$, \ $Z_i = 1$ and $\delta = Z_d n_{d,
0}/(Z_i n_{i, 0}) = 0.25$, implying $\mu = 0.75$ for negatively
charged dust (and $\mu = 1.25$ for positively dust charge; cf. the
definitions above). The (normalized) wavenumber $k$, modulation
angle ($0 \le \theta \le \pi$) and the non-thermality parameters
($0 \le a_{e, i} < 1$) will be allowed to vary according to the
focus of our investigation. The negative dust charge case ($s =
-1$) is implied, unless otherwise mentioned.

\subsection{Parallel modulation}

In the special case of the \emph{parallel} modulation ($\theta = 0$),
the analytical behaviour of $P$ and $Q$ was briefly studied in
Ref. \cite{IKPKSEPS}. See that $P = - 3 \omega^5/(2 k^4) < 0$ for
$\theta = 0$ so that one only has to study the sign of $Q$.  For
$k\ll 1$ one then obtains the approximate expression
[\cite{IKPKSEPS}]: $P \approx - 3 k/2$ and $Q \approx (3 + 2 s
\alpha)^2/(12 k)
> 0$, which prescribes stability (and dark/grey type
envelope excitations -- \emph{voids}) for long wavelengths. For
shorter wavelengths, i.e. for $k
> k_{cr}$ (where $k_{cr}$ is some critical wavenumber, to be determined
numerically), $Q$ changes sign (becomes negative), and the wave
may become unstable.

In addition to the above results, an increase in the non-thermal ion 
population (i.e. in the value of $a_i$) is seen to favor instability at lower values of
$k$: 
see that the black region shrinks - for $\theta = 0$ - when passing from Fig. \ref{figure4}a to Fig. \ref{figure4}b. 
Large wavelengths always remain
\emph{stable} and may only propagate as \emph{dark-}type
excitations. 
On the other hand, the effect of an increase in the non-thermal 
\emph{electron} population (i.e. in the value of $a_e$), while keeping ions Maxwellian, has a similiar, yet not 
so dramatic effect; cf. Fig. \ref{figure4}c, for $\theta = 0$. 
We note that the effect of non-thermality (of ions, in particular) on 
the parallel modulation of a dusty plasma with 
\emph{positively} charged dust is qualitatively similar; 
compare e.g. Fig. \ref{figure4}b to Fig. \ref{figure7}b (for $\theta = 0$).

\subsection{Obliqueness effects}

Let us now consider an arbitrary value of $\theta$. First,
we note that, for small wavenumber $k$, one obtains:  $P \approx
\sin^2\theta/(2 k) > 0$ and $Q \approx -(3 + 2 s \alpha)^2/(12 k)
< 0$, so that stability (and the existence of dark/grey type
excitations) is, again, ensured for long wavelengths $\lambda \gg
\lambda_D$.

We proceed by a numerical investigation of the sign of the
coefficient product $P Q$ in the $(k, \theta)$ plane; see Fig.
\ref{figure4}. The profile thus obtained is qualitatively
reminiscent of the results in Ref. \cite{IKPKSPS}: the critical
wavenumber, say $k_{cr}$, beyond which $P Q$ changes sign,
decreases as the modulation obliqueness angle $\theta$ increases
from zero (parralel modulation) to - roughly - 0.5 rad (i.e.
approximately $30°$), implying a destabilization of the DAW
amplitude for lower wave number (higher wavelength) values. Now,
above 0.5 rad (i.e. $60°$), approximately, and up to $\pi/2$ rad
(i.e. $90°$, transverse modulation) the wave remains stable.

A question naturally arises at this stage: how does the
forementioned stability profile depend on the non thermal
character of the background? The ion and electron non-thermality
effects are separately treated in the following paragraphs.

\subsection{Ion non-thermality effects}

Let us vary the value of the ion non-thermality parameter $a_i$.
Passing to a finite value of $a_i = 0.2$ results in a
\emph{narrower} stability region for small angles (i.e. $k_{cr}$,
at which $PQ(k_{cr}) = 0$, decreases with $a_i$, for small
$\theta$) -- see Fig. \ref{figure4}b -- while the influence on the
behaviour for larger angles is less important.
Therefore, a small non Maxwellian ion population (for $a_i$ less than 0.2, roughly) seems to slightly enhance 
modulational instability (i.e. destroy stability) at lower wavenumber values, at least for a small to moderate modulation obliqueness. Slightly higher values of $a_i$ rather favor instability even further. 
Finally, the picture is reversed for very high $a_i$ (above, say, 0.5), 
where the wave is slightly stabilized by increasing $a_i$ (at least for moderate values of $\theta$). 
These effects may be
depicted by keeping the value of $\theta$ fixed, while varying the
value of $a_i$; see Figs. \ref{figure5}. The influence of non-thermality 
on \emph{strongly} oblique DAW modulation is tacitly negative: stability is destroyed by increasing $a_i$ (see e.g. Fig. \ref{figure5}c, where $\theta =\pi/4$). 
The transverse DAW modulation case (i.e. for $\theta =\pi/2$) remains 
globally stable, so the corresponding (black) diagram was omitted in 
Fig. \ref{figure5}.

\subsection{Electron non-thermality effects}

In analogy to the previous paragraph, we may now vary the value of
the electron non-thermality parameter $a_e$. The effect is
qualitatively similar, yet far less dramatic; see Fig.
\ref{figure6}. This is physically expected, since the influence of
the background ions on the inertial dust grains is more important
than that of the lighter electrons.

\subsection{Positive dust}

The analogous investigation in the positive dust case ($s=+1$;
$\mu = 1.25$ everywhere below) reveals a qualitatively different
picture. Again, the addition of positive dust is seen to
result in a much \emph{narrower} stability region; 
compare Figs. \ref{figure4} and \ref{figure7}. We see that positive dust 
slightly favors instability, with respect to negative dust; 
cf. the discussion in Ref. \cite{IKPKSPS}.

As far as the influence of ion non-thermality on stability is
concerned, the qualitative effect seems to be the \emph{opposite}
as in the preceding paragraphs. For a given value of the angle $\theta$,
increasing the ion ``non-Maxwellian'' parameter $a_i$ seems to
result in a \emph{more unstable} configuration, i.e. in a narrower 
black region in Fig. \ref{figure8}, 
for $\theta=0$; this is true for values of $a_i$ up to, say, 0.5, while 
reaching higher ones bears the opposite effect (see the upper half of 
Fig. \ref{figure8}a).
For higher
$\theta$, the result is similar, yet less dramatic; see Figs
\ref{figure8}b, c.  
Again (as in the positive dust charge case), 
the transverse DAW modulation case (i.e. for $\theta =\pi/2$) remains 
globally stable, so the corresponding (completely black) diagram was omitted in Fig. \ref{figure8}. 

Finally, taking into account electron non-thermality, i.e.
increasing the electron parameter $a_e$ from zero to a finite
value yields a similar effect, yet much less important effect, for
practically all values of $\theta$ and $k$.

In conclusion, the modulational instability of DAWs propagating in a dusty plasma
with \emph{positively} charged dust
grains seems to be enhanced by the appearance of a non-thermal
background component.

\section{Conclusions} We have investigated the amplitude modulation of
dust acoustic waves  in the presence of a
 non-thermal (non-Maxwellian) ion and/or electron background,
 focusing on the influence of
 the latter on the stability profile, and also on the conditions for the
occurrence of envelope excitations.

Summarizing our results, we find that the presence of 
non-Maxwellian electron an/or ion distribution(s) is expected to yield a
\emph{destabilizing} effect on the DAW propagation in a plasma with
negatively charged dust grains (slightly favoring dark- rather than
bright-type envelope wave packets), in particular for a moderate
obliqueness in amplitude perturbation. 
This effect is (qualitatively similar but)
stronger for nonthermal ions rather than electrons. In the
presence of a transverse modulation, the wave's stability profile
remains intact.
The modulational instability of the DAWs propagating in a
dusty plasma with positively
charged dust grains ($q_d > 0$) also seems to be enhanced by the
appearance of a nonthermal electron and ion components. Again, this
effect is stronger for nonthermal ions rather than electrons;
also,  transversely modulated DAWs remain stable. 

Finally, the occurrence of bright
(rather than dark) type envelope wave packets is rather favored by deviations of the ion (and electron, though less) population(s) 
from the Maxwellian distribution. 

Therefore, adding a non thermal ion and/or electron component to the plasma
may control (and possibly stabilize) the propagation of dust-acoustic 
envelope solitons, by enhancing energy localization via instability of the wave's amplitude (due to carrier self-modulation). 

\begin{acknowledgments}
This work was supported by the DFG ({\it{Deutscheforschungsgemeischaft}})
 through the Programme 
{\it{SFB591
(Sonderforschungsbereich) -- Universelles Verhalten
gleichgewichtsferner Plasmen: Heizung, Transport und
Strukturbildung}}.

The authors are happy to dedicate this article to Robert Alan
Cairns on the occasion of his 60th birthday.
\end{acknowledgments}

\newpage

\bigskip

\textbf{\Large{Figure captions}}

\bigskip

\bigskip

Figure 1:

The non-thermal distribution $f(v; a)$, as defined by Eq.
(\ref{Cairns-model1}) [scaled by $f(v; 0)=1/\sqrt{2 \pi}$], vs.
the reduced velocity $v/v_{th}$, for $a =$ 0, 0.1, 0.2, 0.3 (from
top to bottom).

\bigskip
\bigskip

Figure 2:

\emph{Bright} type localized envelope modulated wavepackets (for
$P Q > 0$) for two different (arbitrary) sets of parameter
values.

\bigskip

\bigskip

Figure 3:

(a) \emph{Dark} and (b) \emph{grey} type localized envelope
modulated wavepackets (for $P Q < 0$). See that the amplitude
never reaches zero in the latter case.

\bigskip

\bigskip

Figure 4:

The sign of the product $P Q$ is depicted vs. (normalized)
wavenumber $k$ (horizontal axis) and modulation angle $\theta$
(vertical axis), via a color code: black (white) regions denote a
negative (positive) sign of $P Q$, implying stability
(instability), and suggesting dark (bright) type envelope soliton
occurrence. Here, $\mu = 0.75$ (negative dust charge) and: (a)
$a_i = a_e = 0$ (Maxwellian background); (b) $a_i = 0.2$,  $a_e =
0$ (ion non-thermality); (c) $a_i = 0$,  $a_e = 0.2$ (electron
non-thermality). Remaining parameter values as cited in the text.
See that the effect in (c) is rather negligible.

\bigskip

\bigskip

Figure 5:

The sign of the product $P Q$ is depicted vs. (normalized)
wavenumber $k$ (horizontal axis) and ion non-thermality parameter
$a_i$ (vertical axis), for a modulation angle $\theta$ equal to:
(a) 0 (parallel modulation); (b) 0.4; (c) $\pi/4$; the case
$\theta = \pi/2$ -- transverse modulation -- is omitted, since
globally stable. Same color code as in Fig. \ref{figure4}: black
(white) regions denote a negative (positive) sign of $P Q$. Here,
$\mu = 0.75$ (negative dust charge) and $a_e = 0$ (Maxwellian
electrons). Remaining parameter values as cited in the text.

\bigskip

\bigskip

Figure 6:

Similar to the preceding Figure, but for non-thermal electrons and
Maxwellian ions: the sign of the product $P Q$ is depicted vs.
(normalized) wavenumber $k$ (horizontal axis) and electron
non-thermality parameter $a_e$ (vertical axis), for a modulation
angle $\theta$ equal to: (a) 0 (parallel modulation); (b) 0.4; (c)
$\pi/4$; the case $\theta = \pi/2$ -- transverse modulation -- is
omitted, since globally stable. Same color code as in Fig.
\ref{figure4}: black (white) regions denote a negative (positive)
sign of $P Q$. Here, $\mu = 0.75$ (negative dust charge) and $a_i
= 0$ (Maxwellian ions). Remaining parameter values are cited in the
text.

\bigskip

\bigskip

Figure 7:

Similar to Figure \ref{figure4}, but for a \emph{positive} dust
charge; here, $\mu = 1.25$. The sign of the product $P Q$ is
depicted vs. (normalized) wavenumber $k$ (horizontal axis) and
modulation angle $\theta$ (vertical axis), for: (b) $a_i = 0.2$,
$a_e = 0$ (ion non-thermality); (c) $a_i = 0$,  $a_e = 0.2$
(electron non-thermality). Remaining parameter values as in Fig.
\ref{figure4}. See that the effect in (c) is not as dramatic as in
(b).

\bigskip

\bigskip

Figure 8:

Similar to Figure \ref{figure5}, but for a \emph{positive} dust
charge; here, $\mu = 1.25$ and $a_e = 0$ (remaining parameter
values as in Fig. \ref{figure5}).  The sign of the product $P Q$
is depicted vs. (normalized) wavenumber $k$ (horizontal axis) and
ion non-thermality parameter $a_i$ (vertical axis), for a
modulation angle $\theta$ equal to: (a) 0 (parallel modulation);
(b) 0.4; (c) $\pi/4$; the case $\theta = \pi/2$ -- transverse
modulation -- is omitted, since globally stable.

\bigskip

\bigskip

Figure 9:

Similar to Figure \ref{figure6}, but for a \emph{positive} dust
charge; here, $\mu = 1.25$ and $a_i = 0$ (remaining parameter
values as in Fig. \ref{figure6}).  The sign of the product $P Q$
is depicted vs. (normalized) wavenumber $k$ (horizontal axis) and
electron non-thermality parameter $a_e$ (vertical axis), for a
modulation angle $\theta$ equal to: (a) 0 (parallel modulation);
(b) 0.4; (c) $\pi/4$; the case $\theta = \pi/2$ -- transverse
modulation -- is omitted, since globally stable. See that the
electron non-thermality effect is less important than its ion
counterpart (cf. Fig. \ref{figure6}).


\newpage

\bigskip

\bigskip

\begin{figure}[htb]
 \centering
 \resizebox{3.5in}{!}{ \includegraphics[]{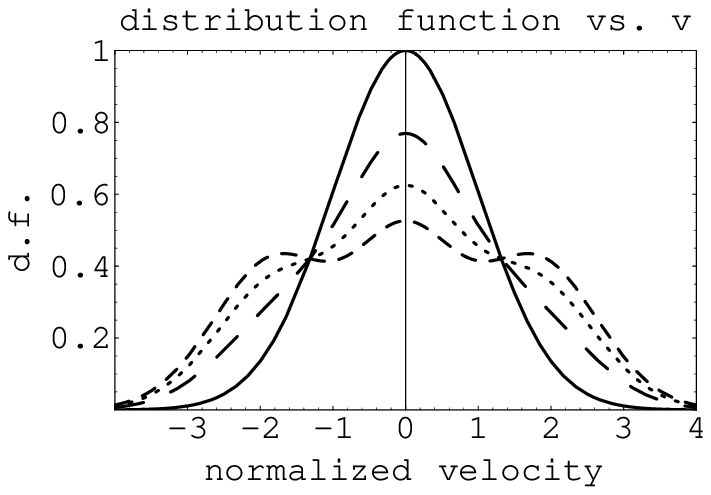} }
 \\
 \vskip 2 cm
\caption{} \label{figure1}
\end{figure}

\newpage

\vspace{4cm}

\begin{figure}[htb]
 \centering
 \resizebox{3.5in}{!}{ \includegraphics[]{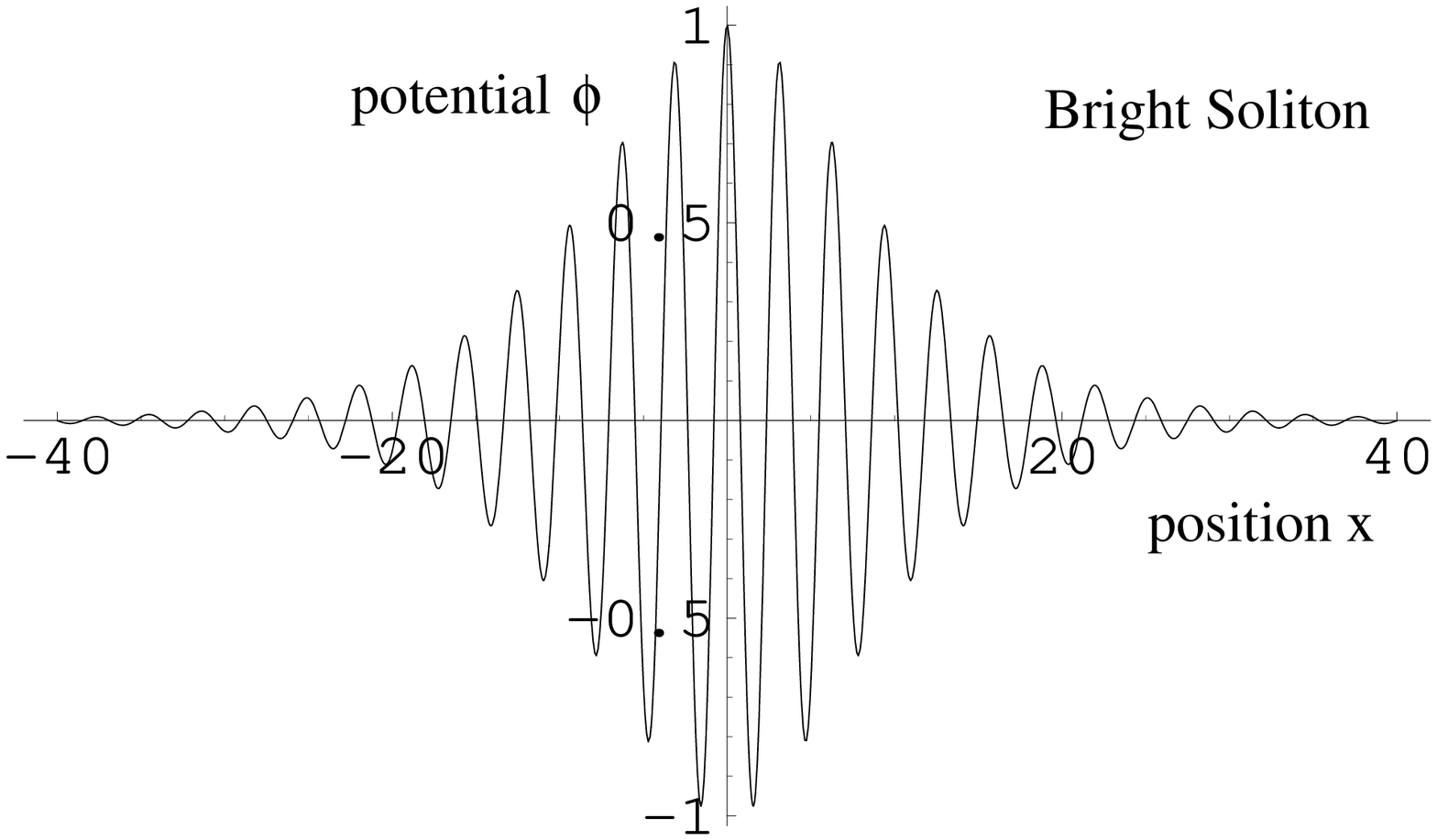} } \\
 \vskip 1 cm {\Large{(a)}} \\
 \vskip 3 cm
 \resizebox{3.5in}{!}{ \includegraphics[]{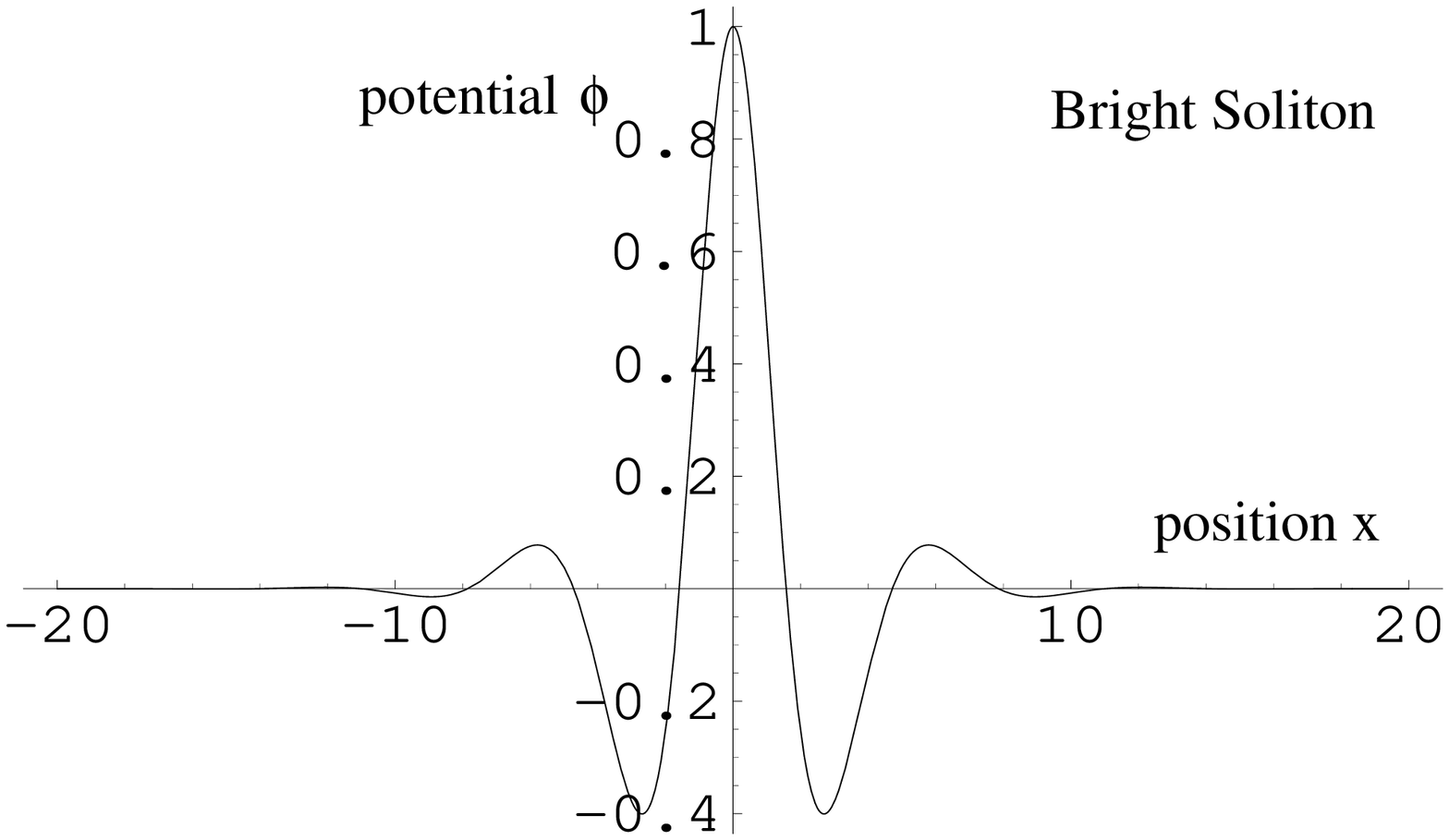}
 } \\
 \vskip 1 cm {\Large{(b)}}
 \\
 \vskip 4 cm
\caption{} \label{figure2}
\end{figure}

\newpage

\vspace{4cm}

\begin{figure}[htb]
 \centering
 \resizebox{3.5in}{!}{ \includegraphics[]{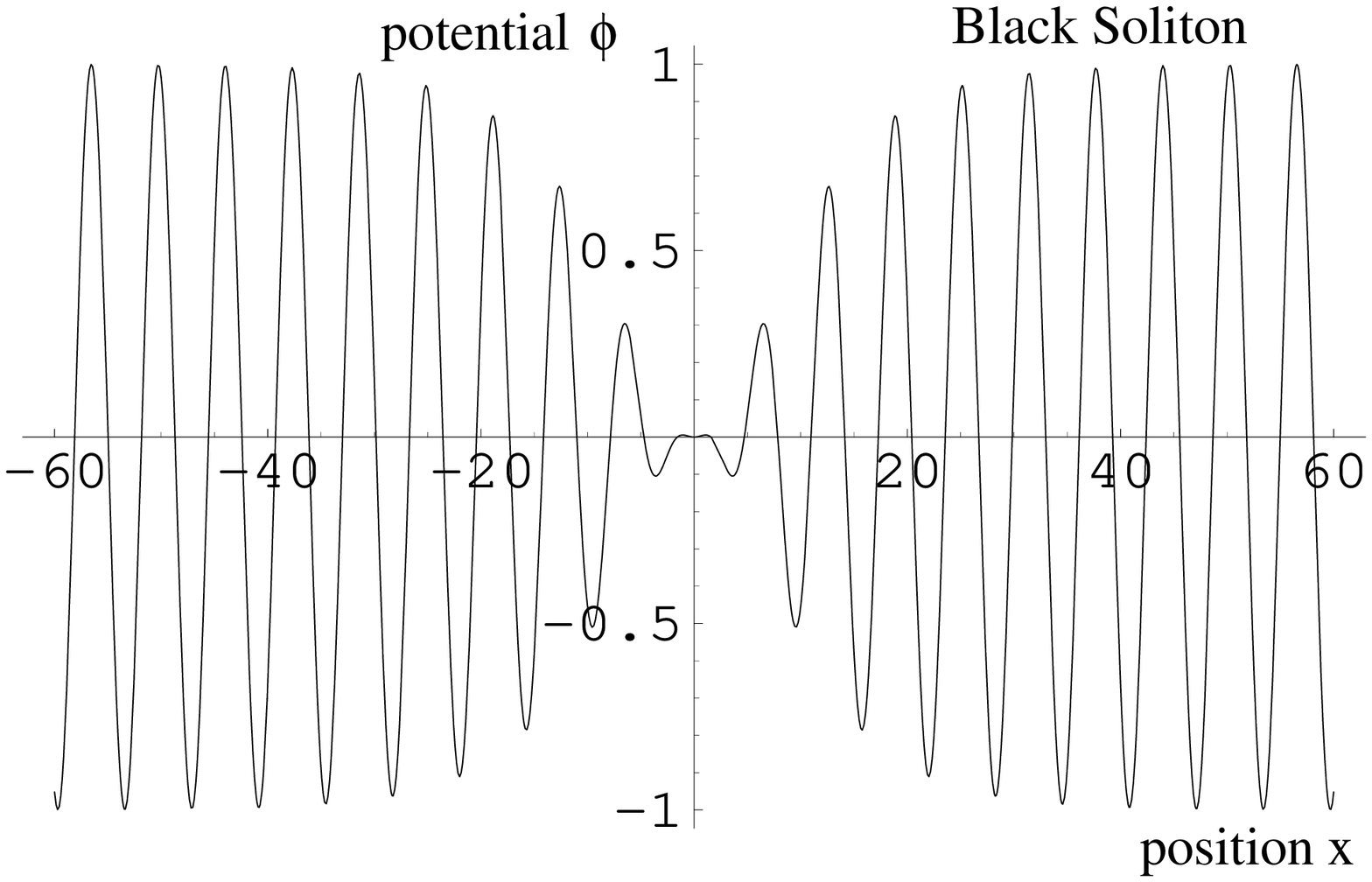} } \\
 \vskip 1 cm {\Large{(a)}} \\
 \vskip 3 cm
 \resizebox{3.5in}{!}{ \includegraphics[]{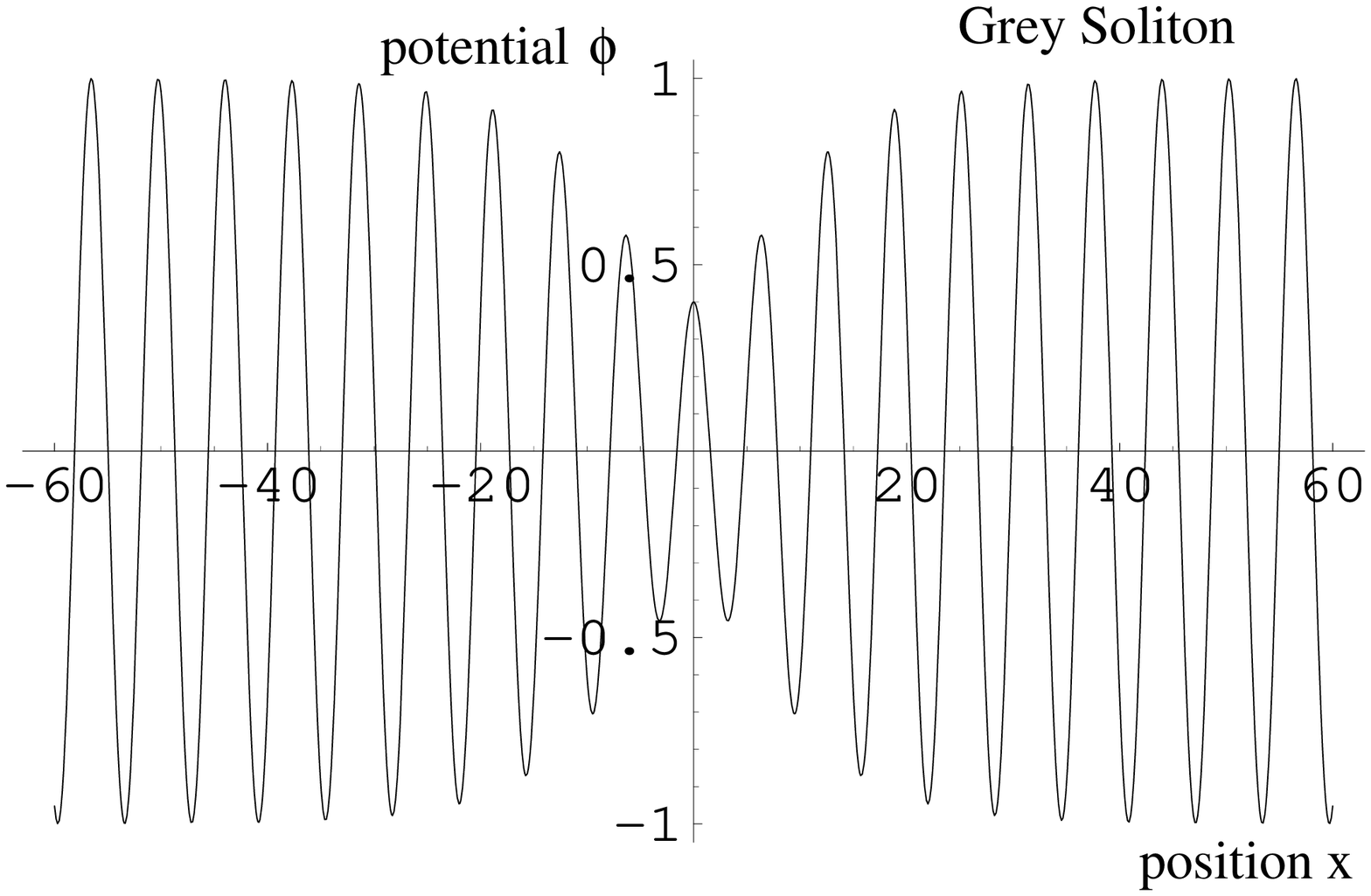}
 } \\
 \vskip 1 cm {\Large{(b)}}
 \\
 \vskip 4 cm
\caption{} \label{figure3}
\end{figure}

\newpage


\begin{figure}[htb]
 \centering
 \resizebox{2.5in}{!}{ \includegraphics[]{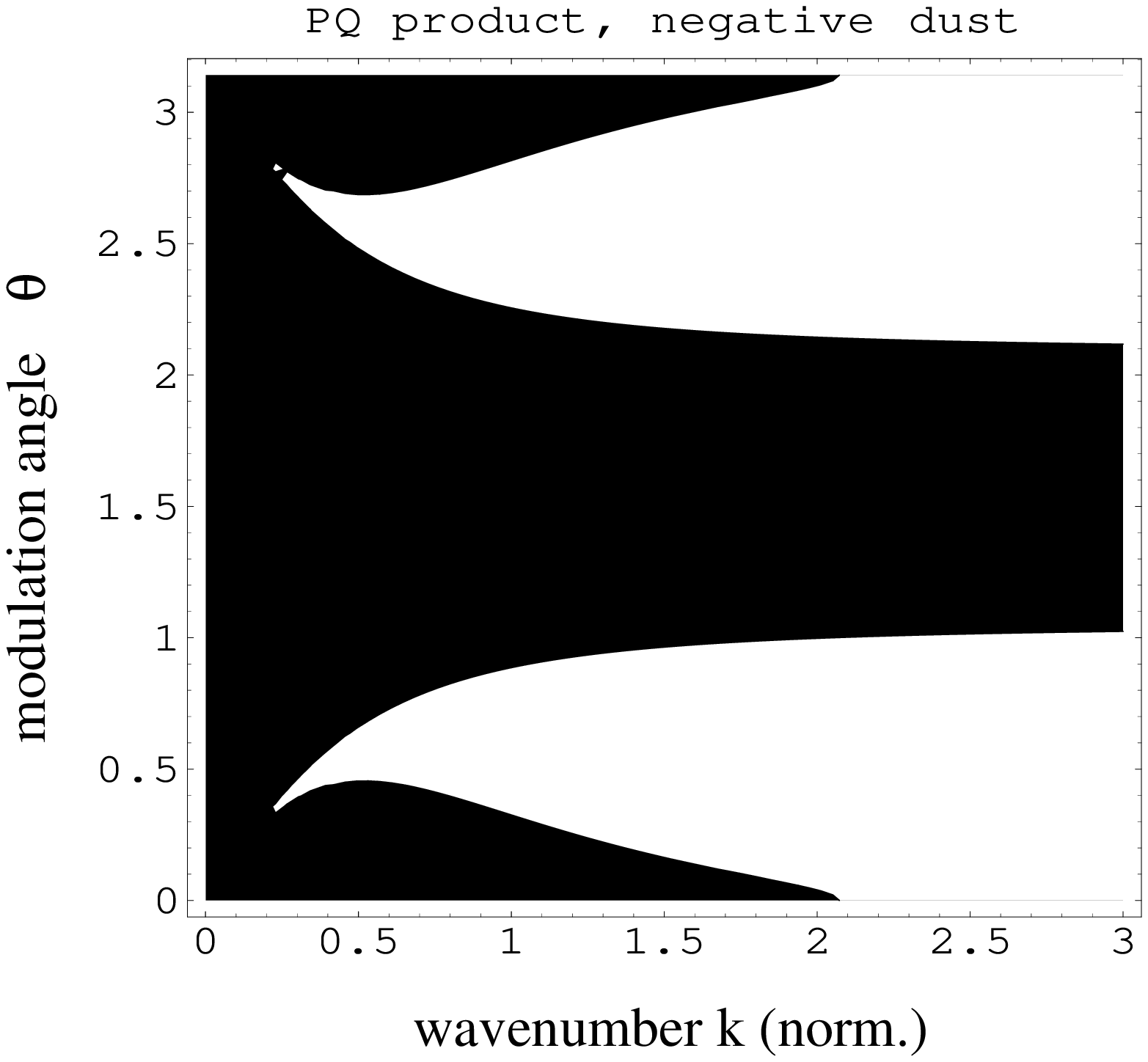} } \\
{\Large{(a)}} \\
\vskip .5 cm
 \resizebox{2.5in}{!}{ \includegraphics[]{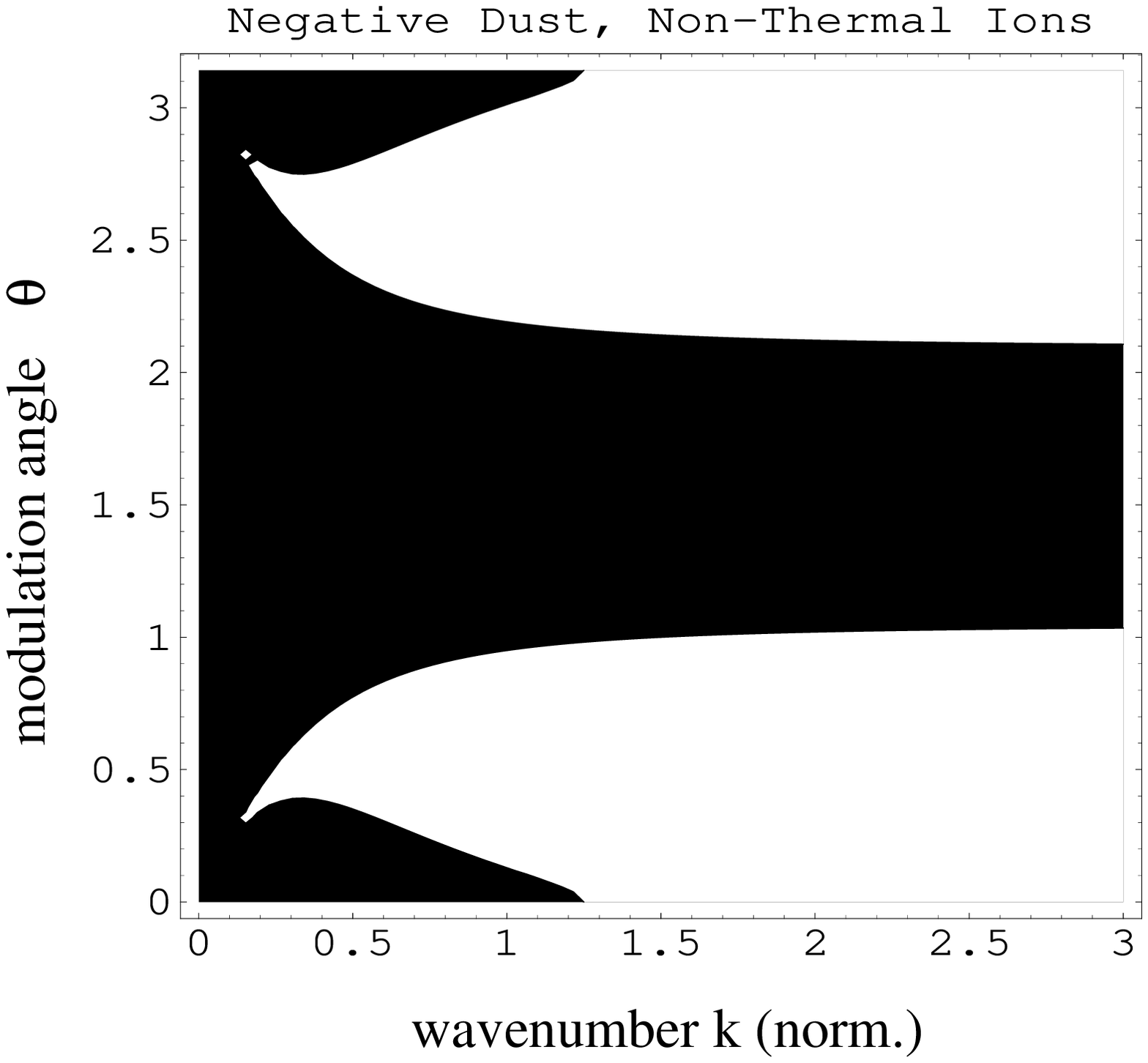}
 } \\
{\Large{(b)}}
 \\
\vskip .5 cm
  \resizebox{2.6in}{!}{ \includegraphics[]{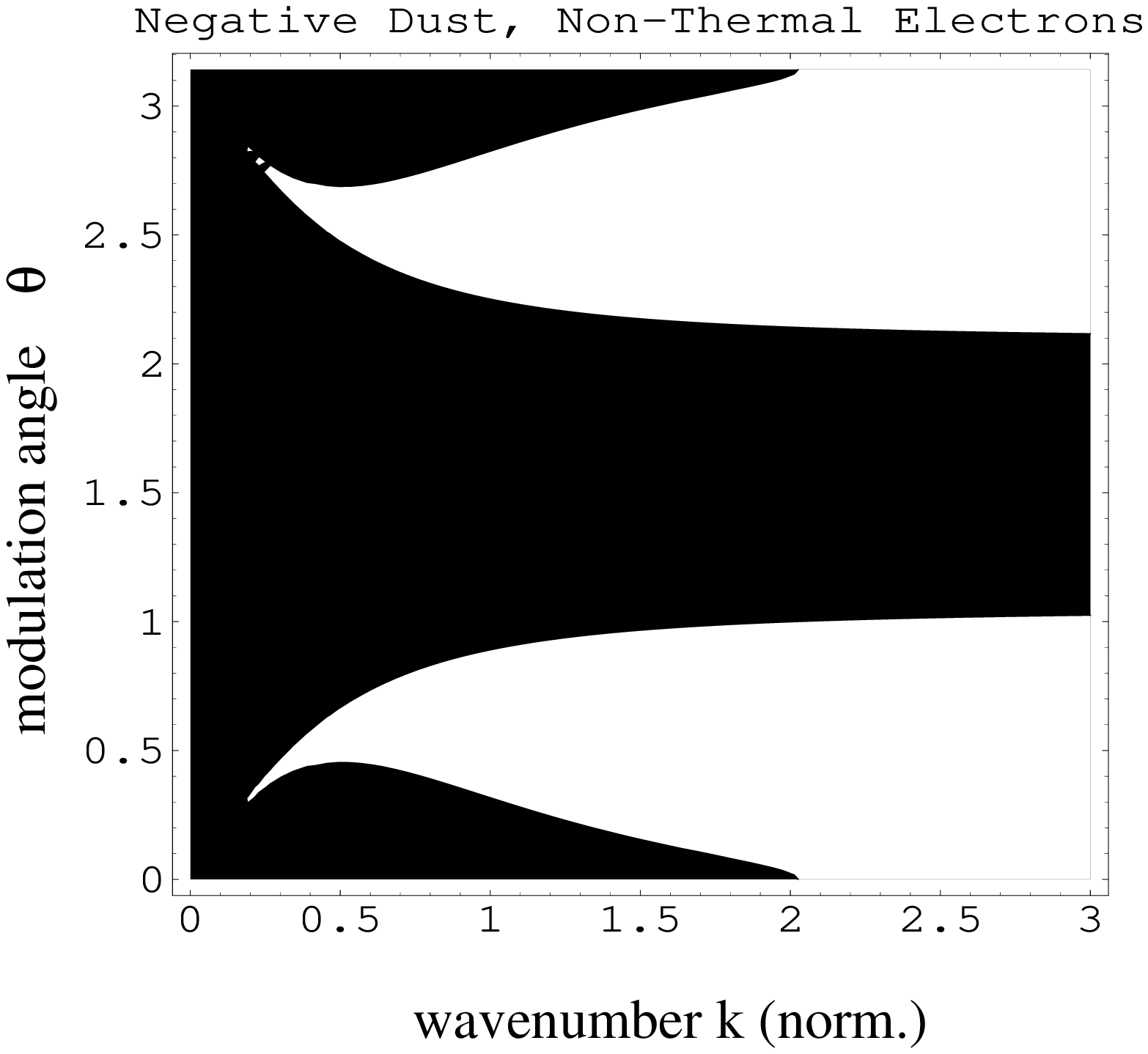}
 } \\
 {\Large{(c)}}
 \\
\vskip .5 cm \caption{} \label{figure4}
\end{figure}

\newpage

\begin{figure}[htb]
 \centering
 \resizebox{2.5in}{!}{ \includegraphics[]{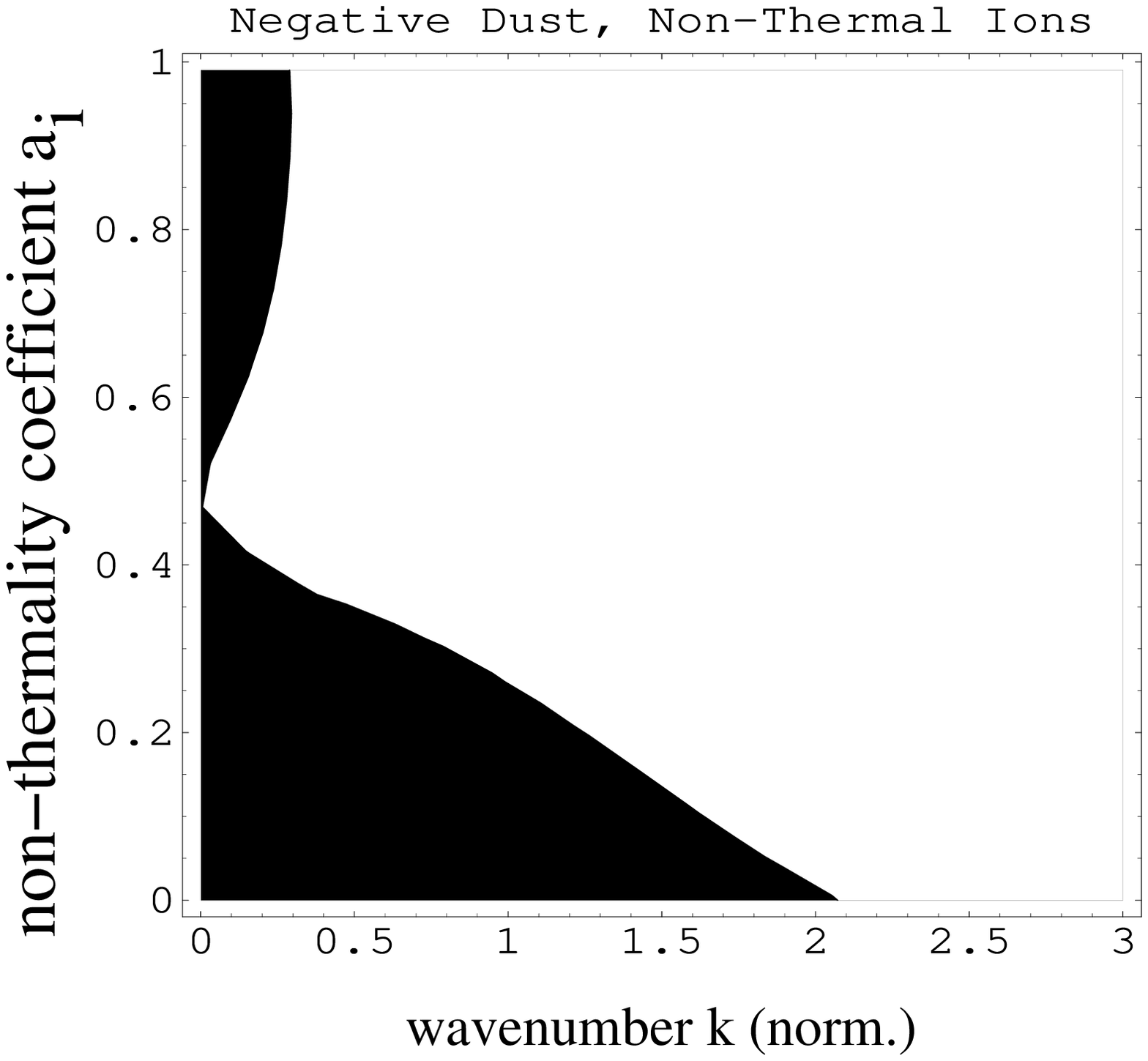} } \\
{\Large{(a)}} \\
\vskip .5 cm
 \resizebox{2.5in}{!}{ \includegraphics[]{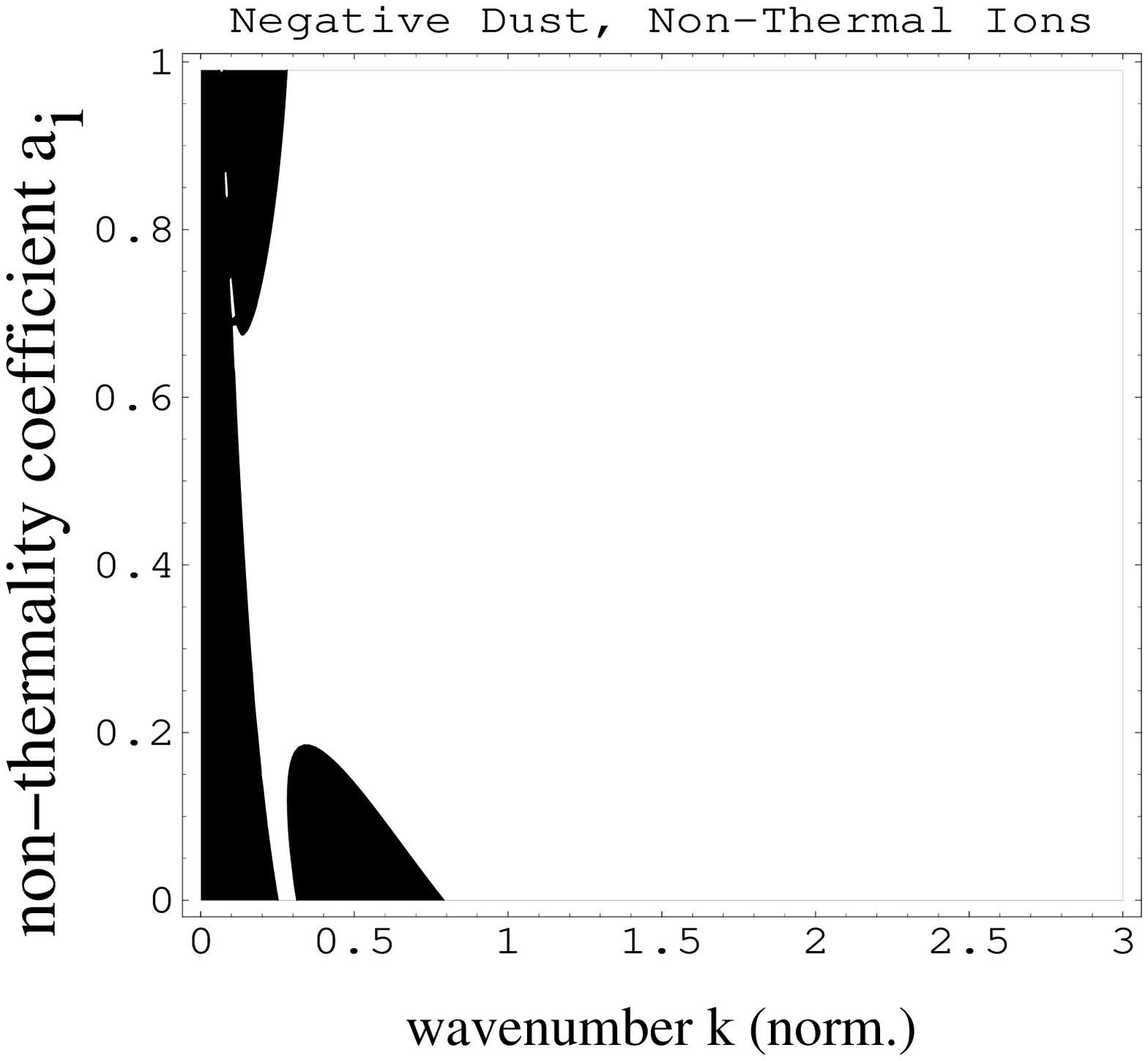}
 } \\
{\Large{(b)}}
 \\
\vskip .5 cm
  \resizebox{2.5in}{!}{ \includegraphics[]{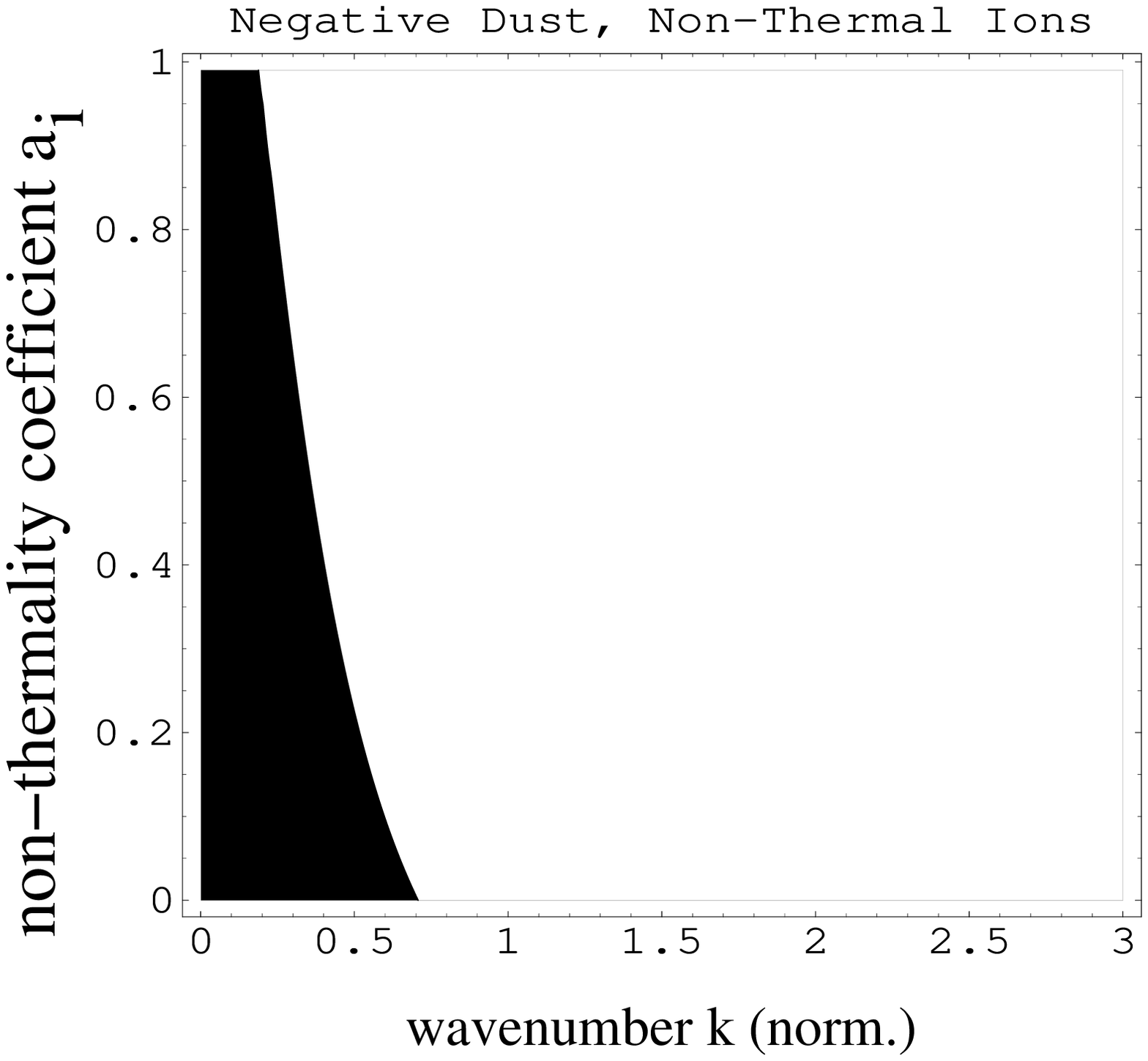}
 } \\
 \vskip .5 cm {\Large{(c)}}
 \\
\caption{} \label{figure5}
\end{figure}

\newpage

\begin{figure}[htb]
 \centering
 \resizebox{2.5in}{!}{ \includegraphics[]{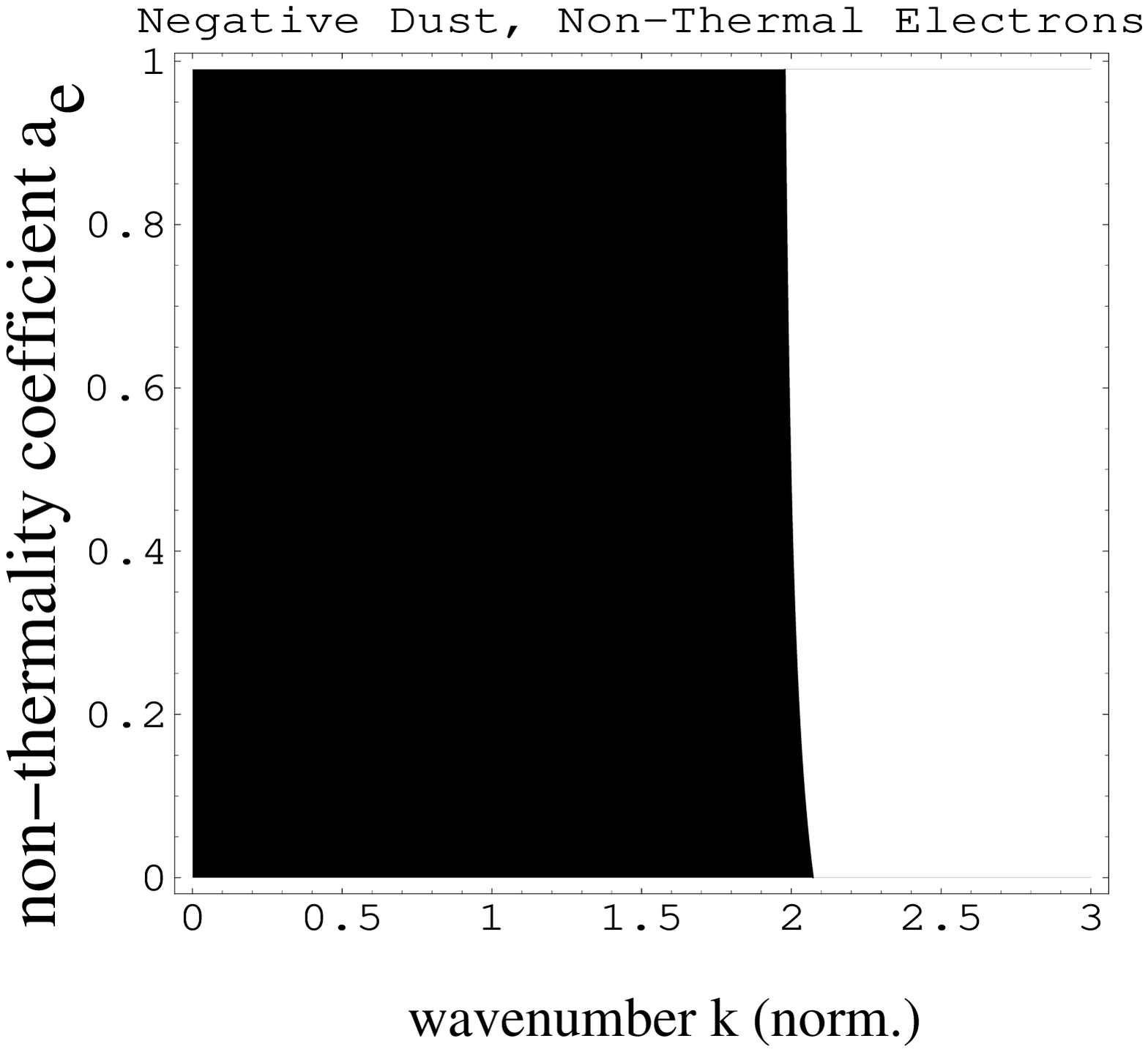} } \\
{\Large{(a)}} \\
\vskip .5 cm
 \resizebox{2.5in}{!}{ \includegraphics[]{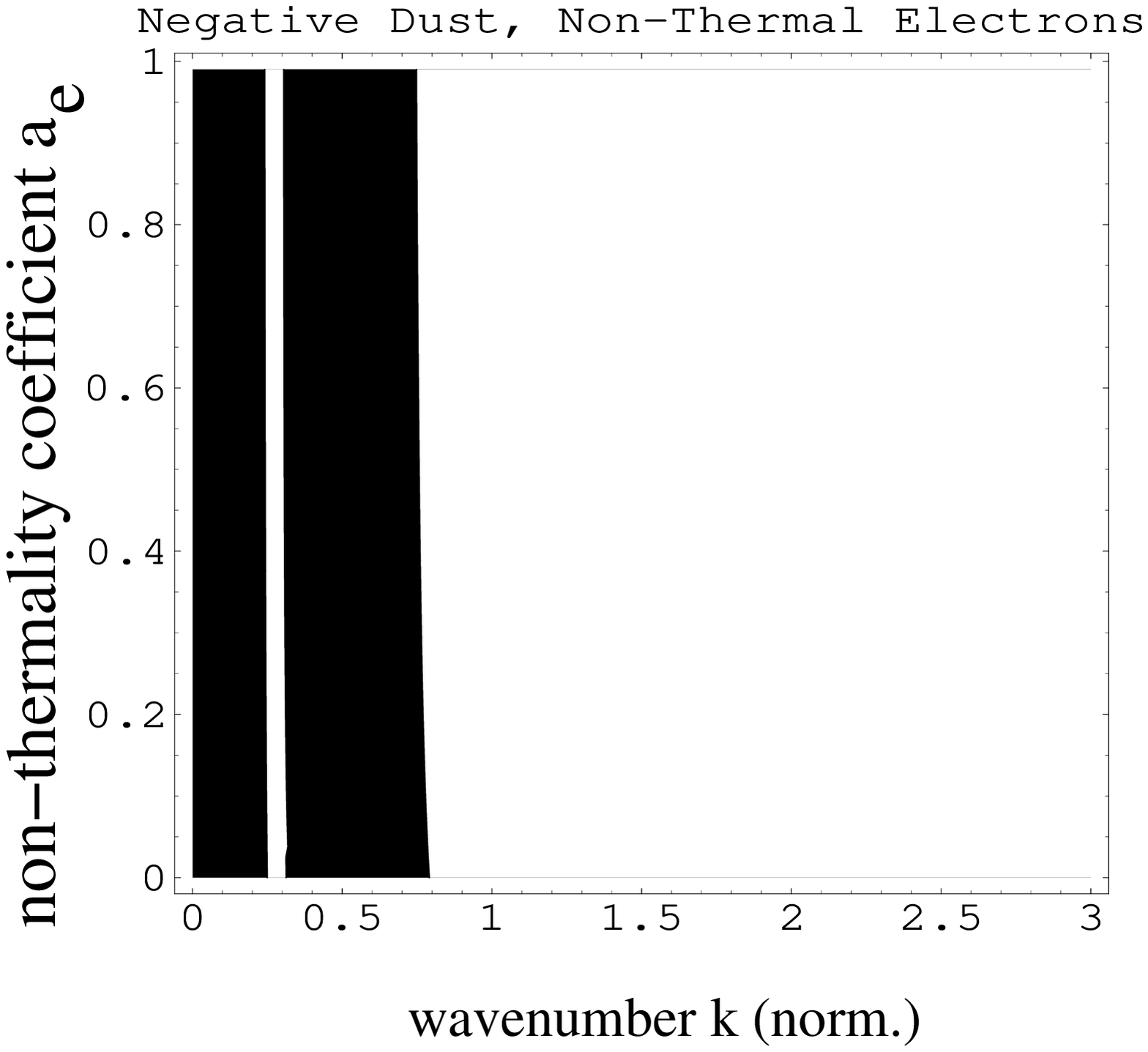}
 } \\
{\Large{(b)}}
 \\
\vskip .5 cm
  \resizebox{2.5in}{!}{ \includegraphics[]{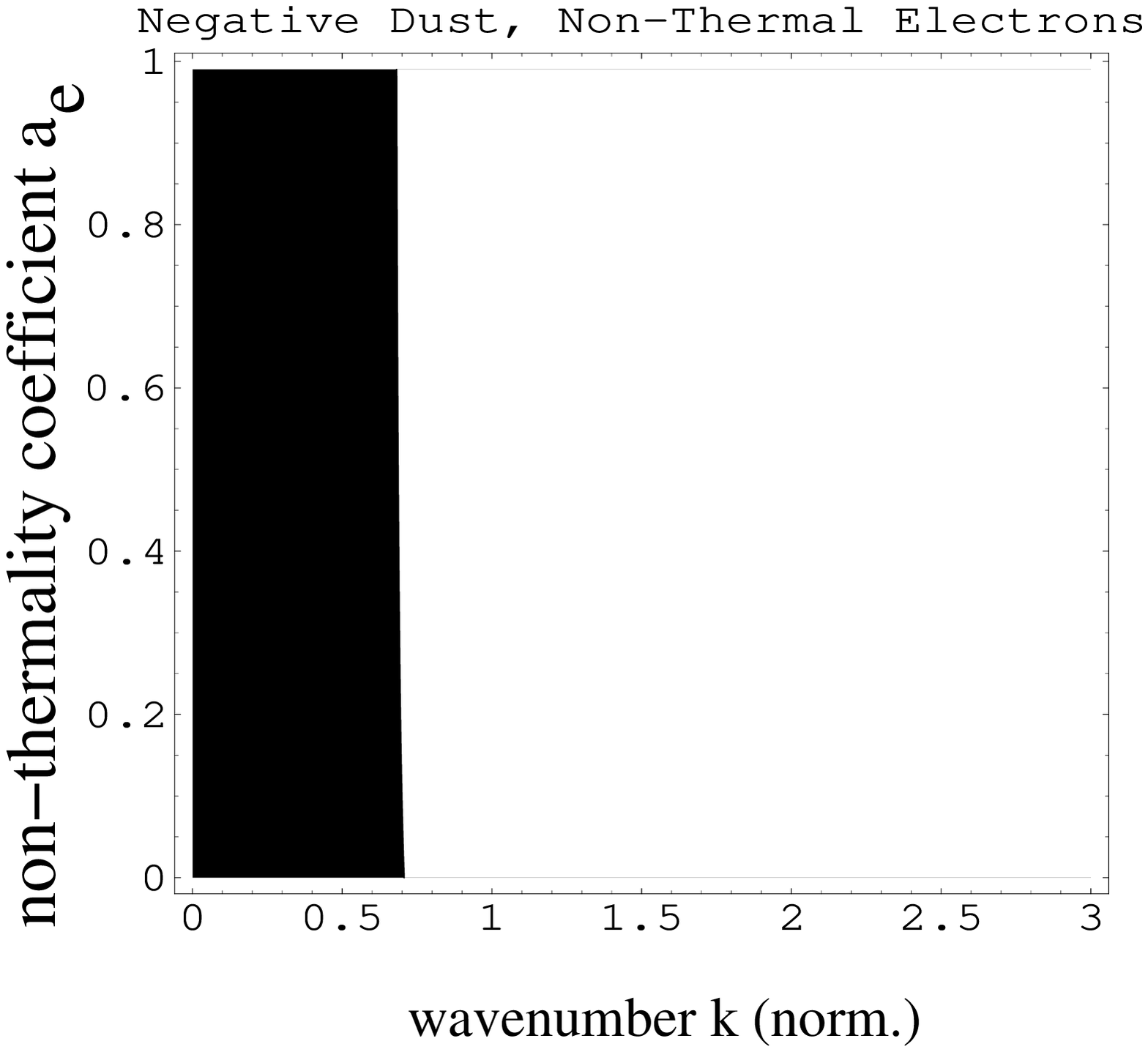}
 } \\
 \vskip .5 cm {\Large{(c)}}
 \\
\caption{} \label{figure6}
\end{figure}

\newpage

\begin{figure}[htb]
 \centering
 \resizebox{2.5in}{!}{ \includegraphics[]{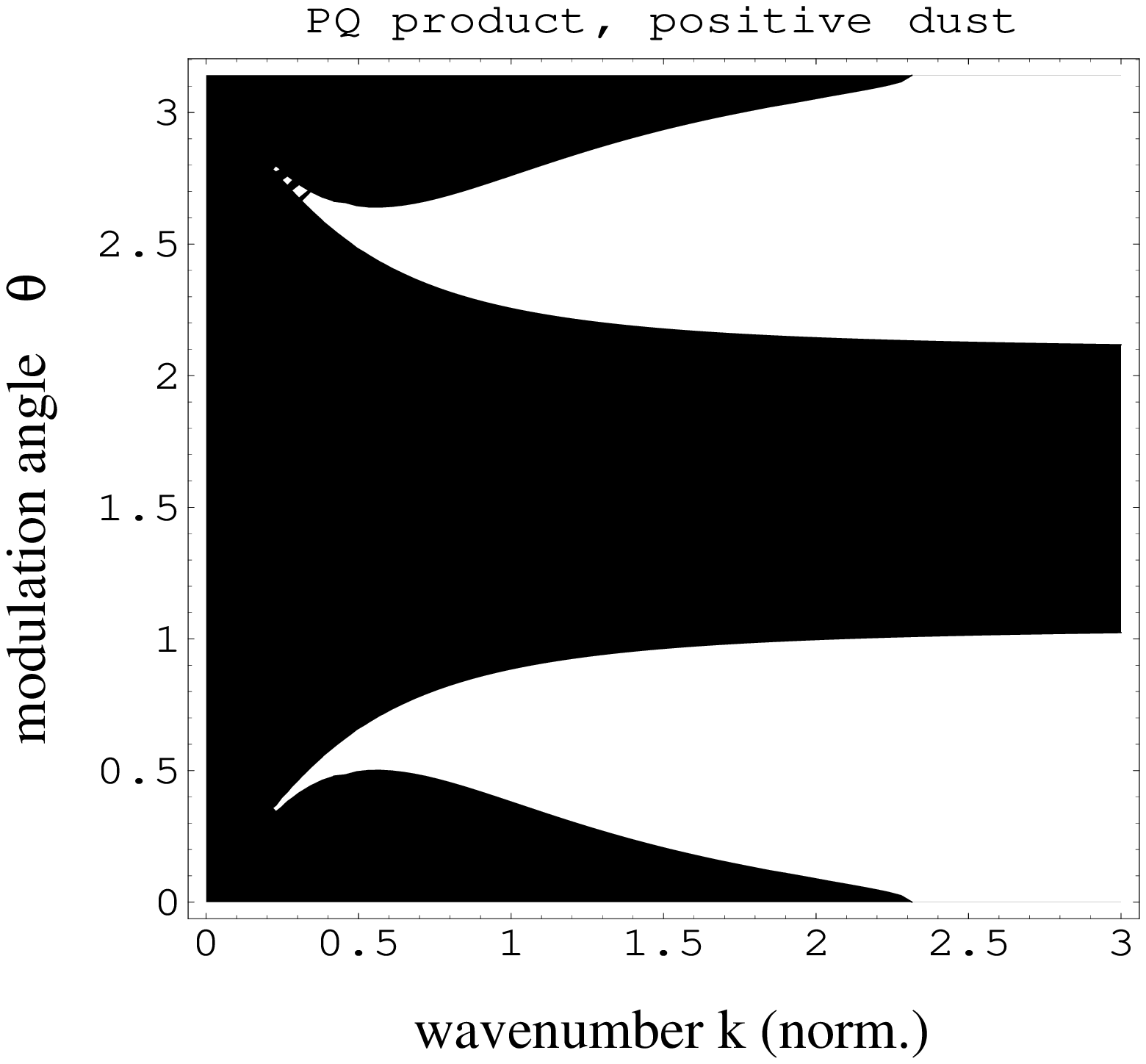} } \\
{\Large{(a)}} \\
\vskip .5 cm
 \resizebox{2.5in}{!}{ \includegraphics[]{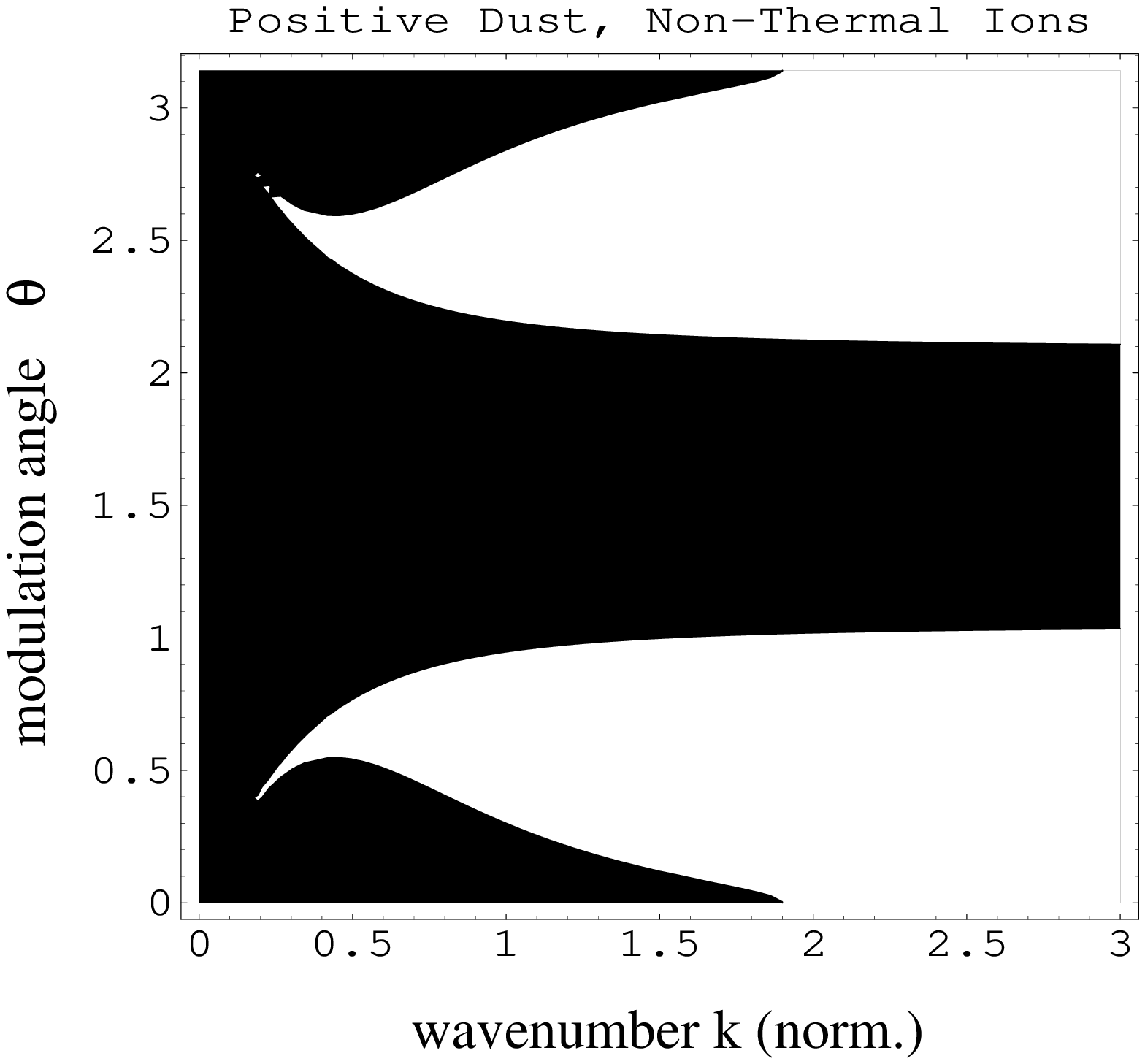}
 } \\
{\Large{(b)}}
 \\
\vskip .5 cm
  \resizebox{2.6in}{!}{ \includegraphics[]{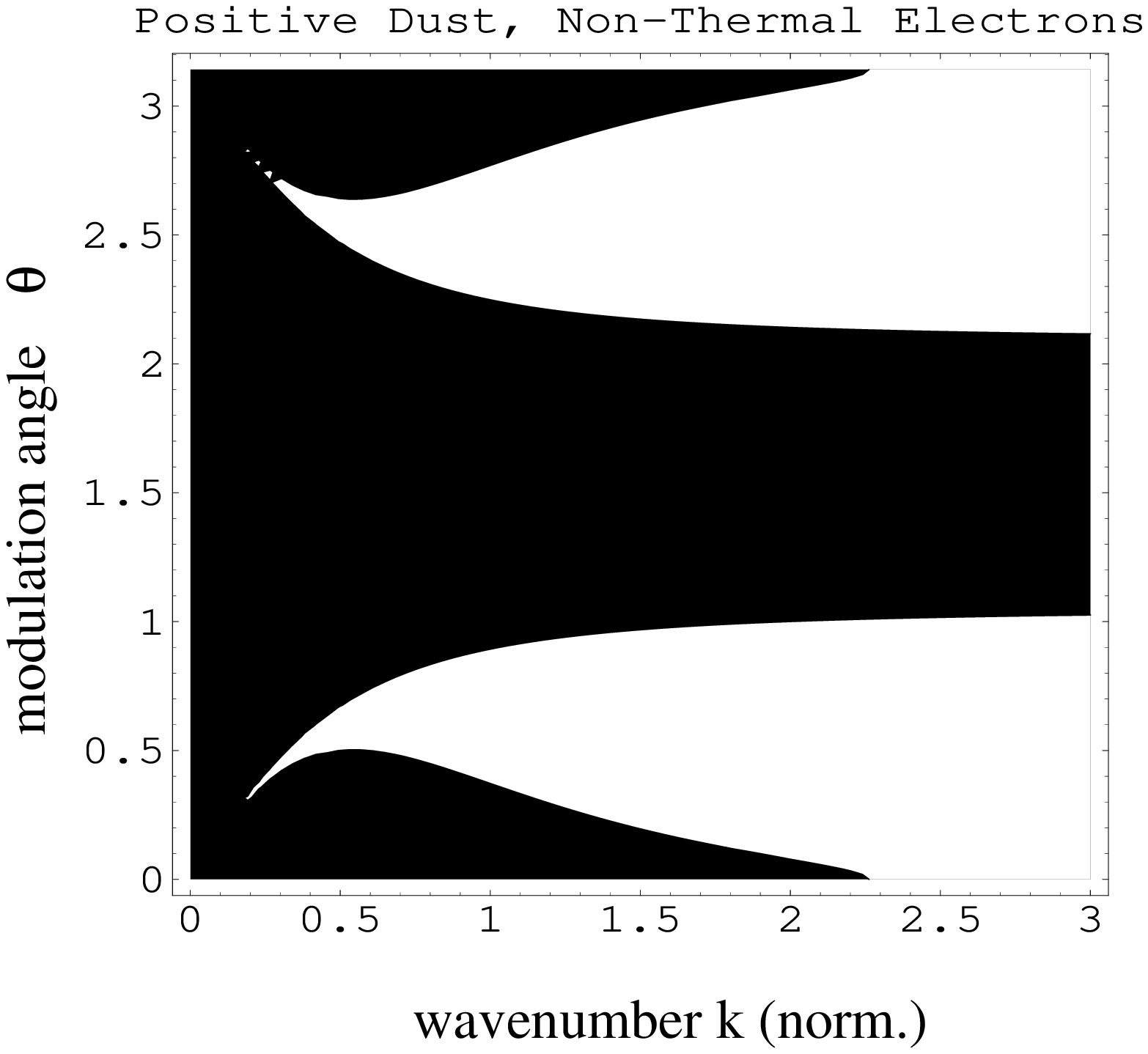}
 } \\
 {\Large{(c)}}
 \\
\vskip .5 cm \caption{} \label{figure7}
\end{figure}

\newpage

\begin{figure}[htb]
 \centering
 \resizebox{2.5in}{!}{ \includegraphics[]{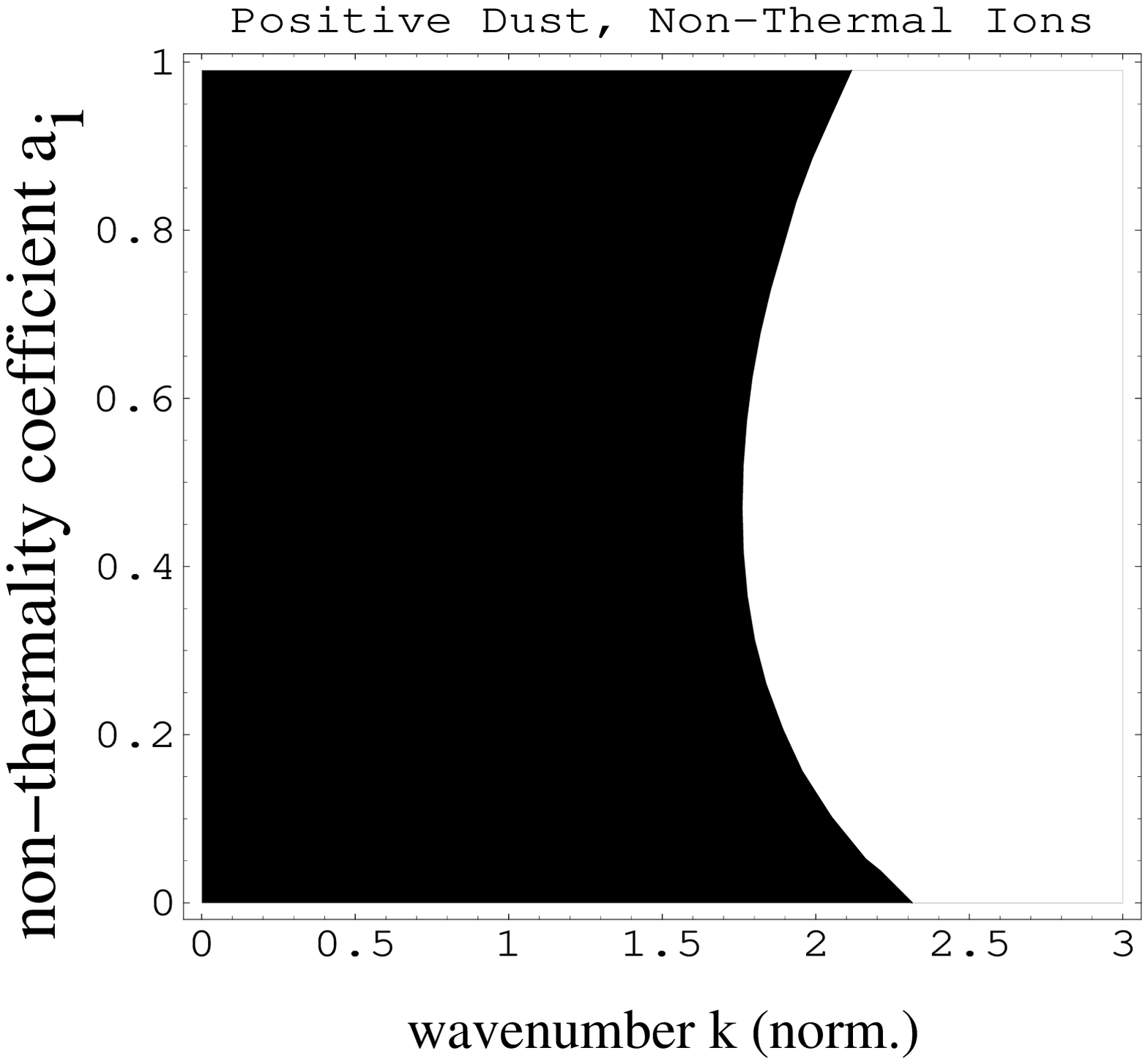} } \\
{\Large{(a)}} \\
\vskip .5 cm
 \resizebox{2.5in}{!}{ \includegraphics[]{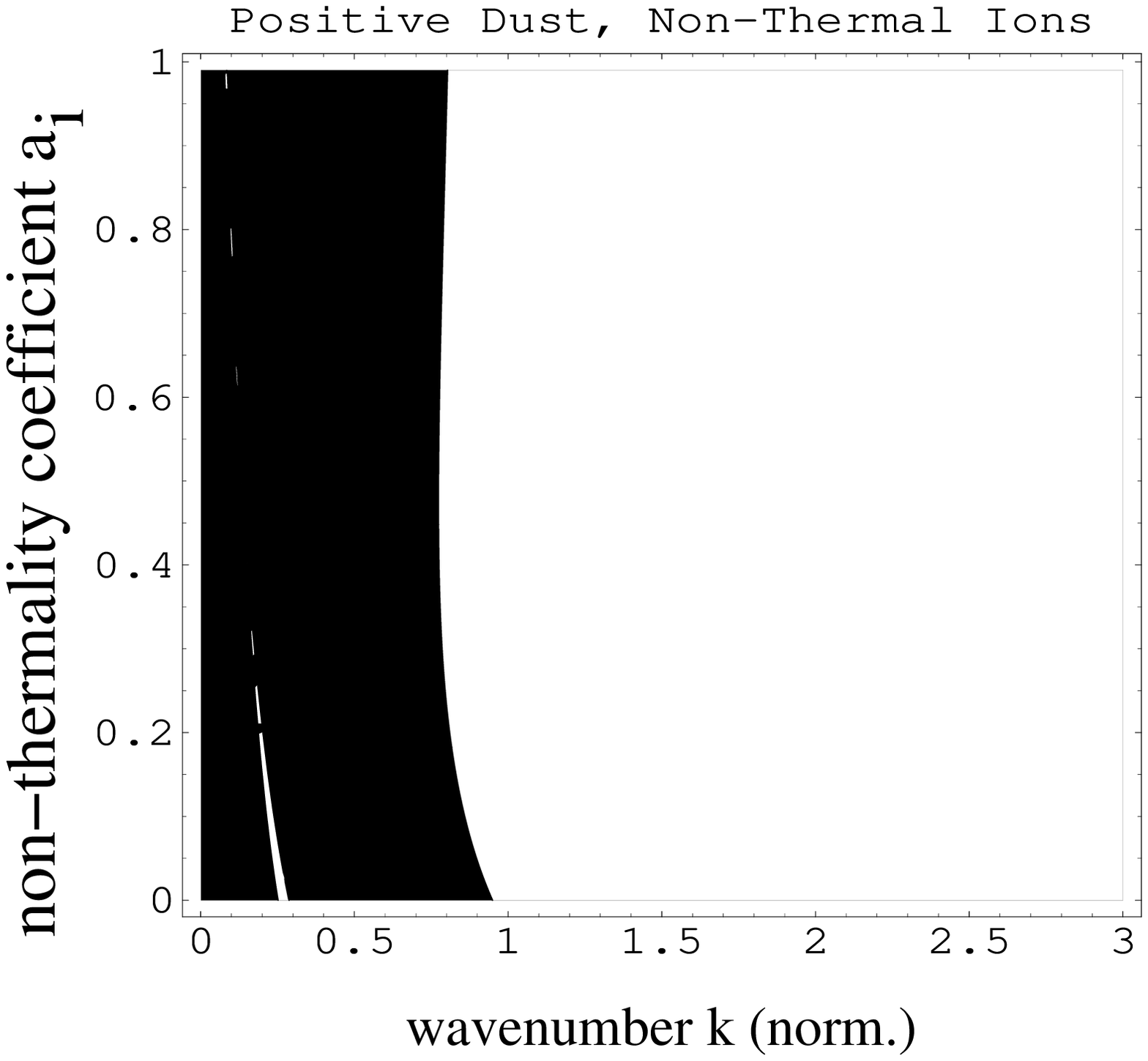}
 } \\
{\Large{(b)}}
 \\
\vskip .5 cm
  \resizebox{2.5in}{!}{ \includegraphics[]{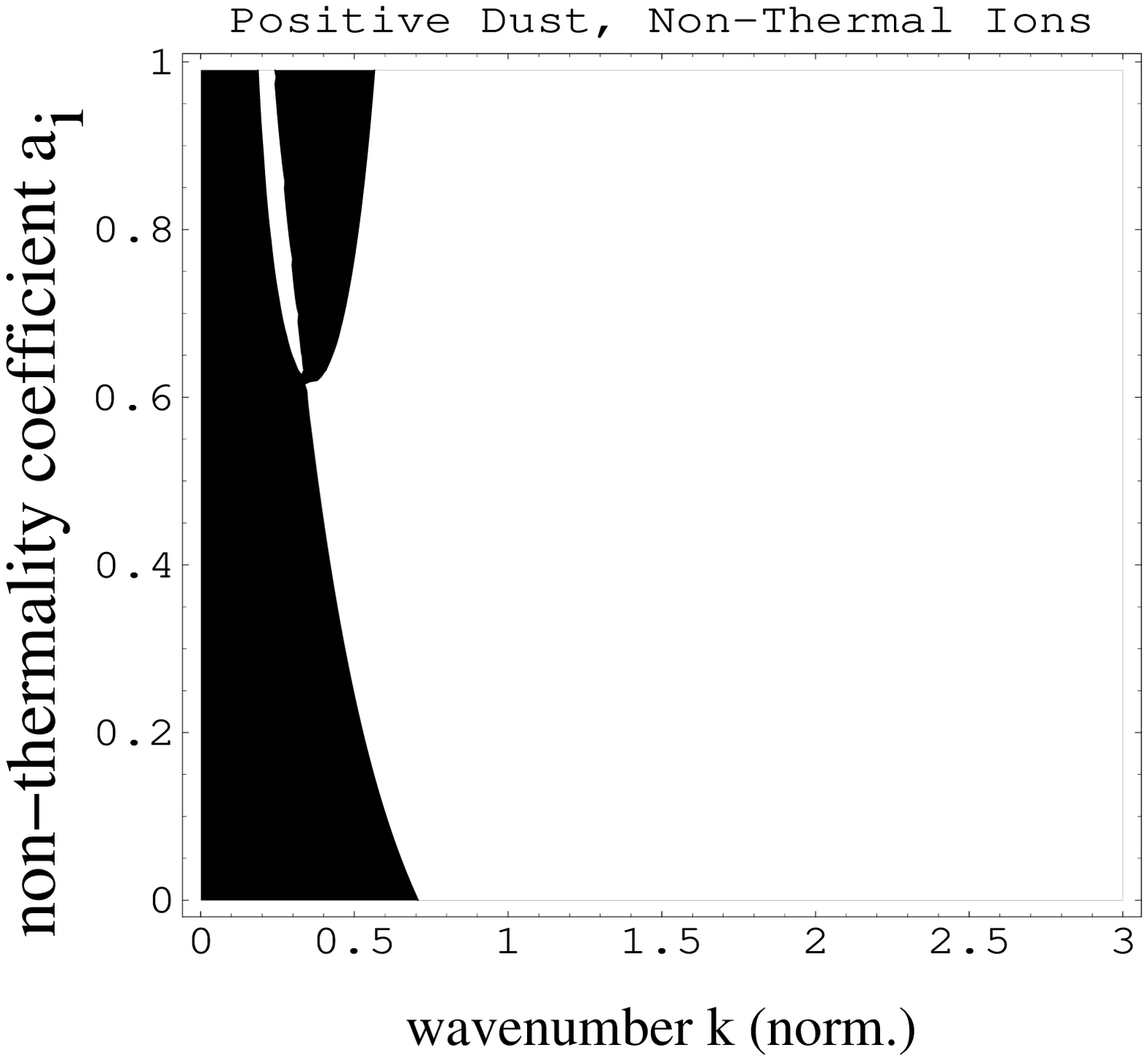}
 } \\
 \vskip .5 cm {\Large{(c)}}
 \\
\caption{} \label{figure8}
\end{figure}

\newpage

\begin{figure}[htb]
 \centering
 \resizebox{2.5in}{!}{ \includegraphics[]{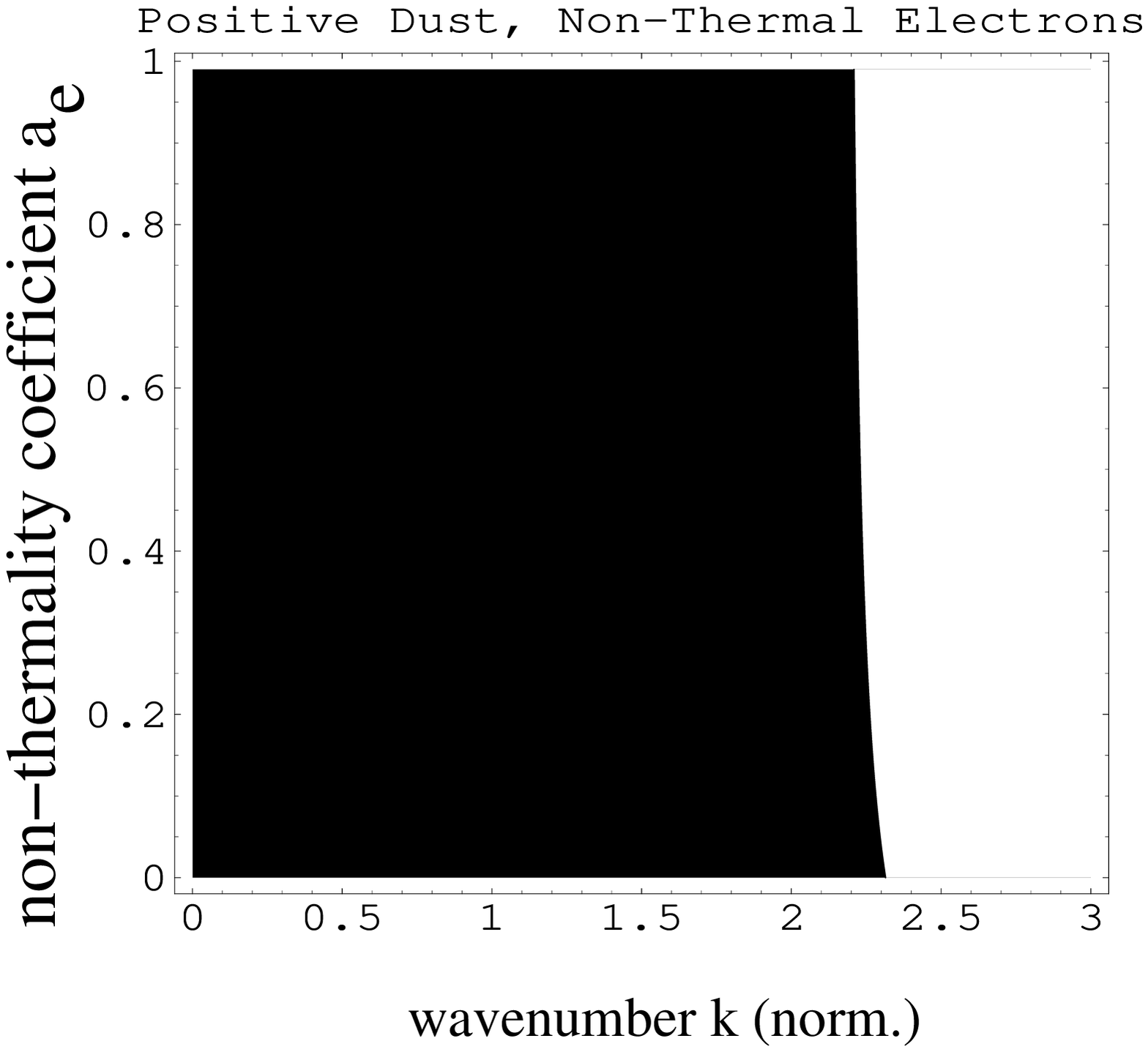} } \\
{\Large{(a)}} \\
\vskip .5 cm
 \resizebox{2.5in}{!}{ \includegraphics[]{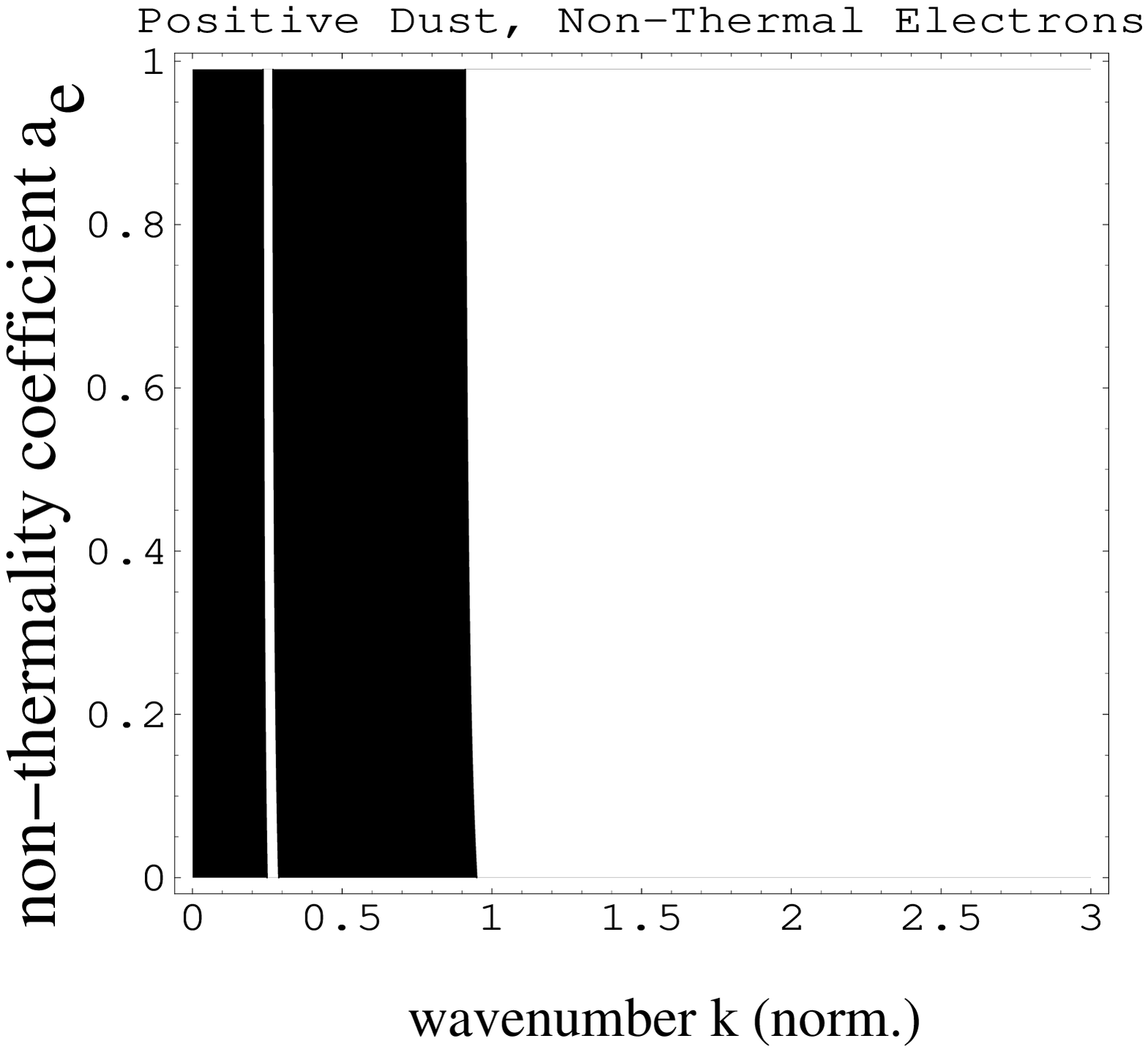}
 } \\
{\Large{(b)}}
 \\
\vskip .5 cm
  \resizebox{2.5in}{!}{ \includegraphics[]{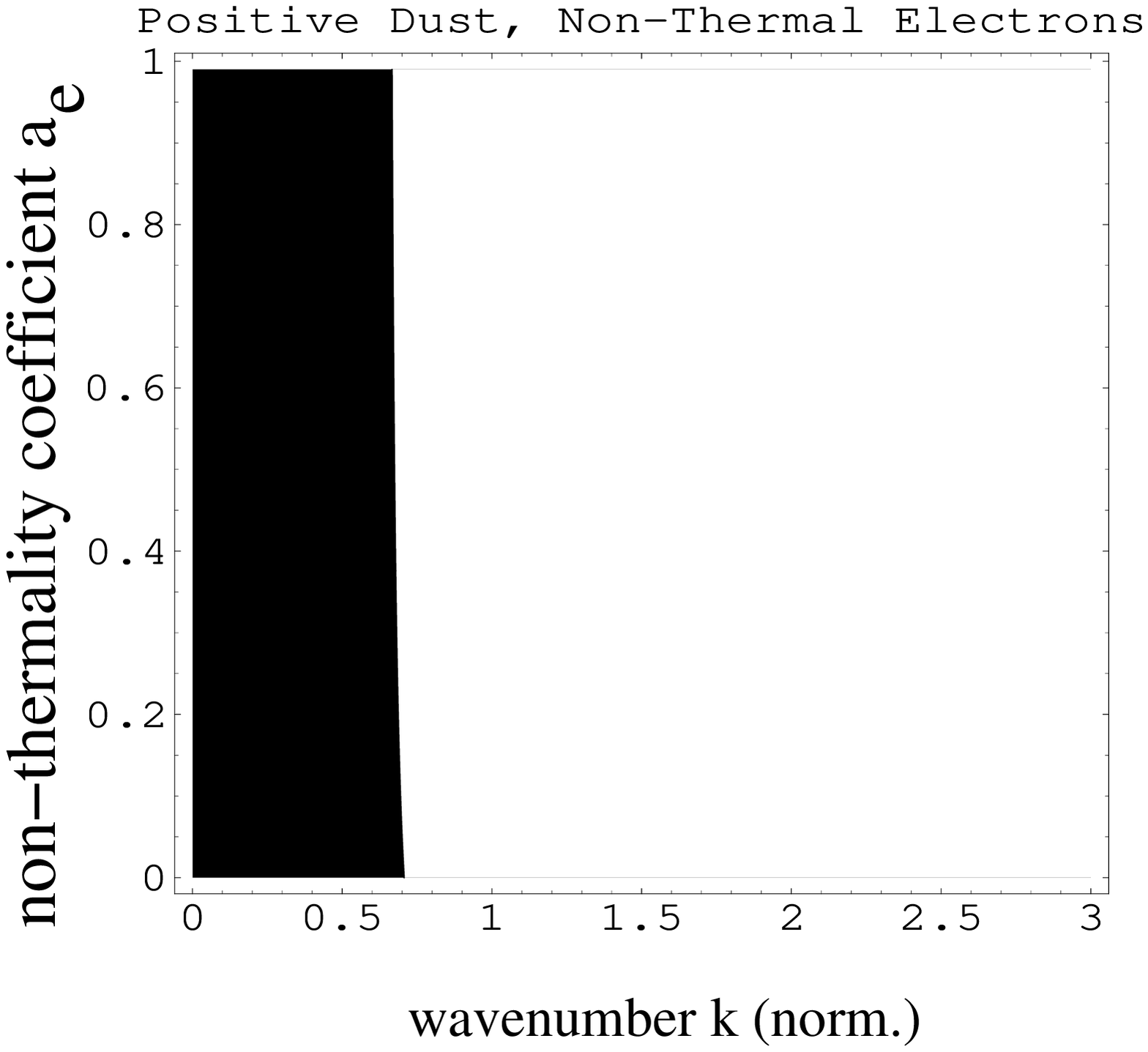}
 } \\
 \vskip .5 cm {\Large{(c)}}
 \\
\caption{} \label{figure9}
\end{figure}


\begin{thereferences}{19}
\bibitem{comment} For rigor, the (reduced) parameter $k$ (i.e. essentially 
$\sim \lambda_{D, eff}/\lambda\equiv k/k_{D, eff}$, where $\lambda$ is the wavelength) 
appearing in the formulae, 
has had to be redefined as $k \rightarrow k \lambda_{D, eff}/\lambda_{D, d}$, in order to be 
represented in the horizontal axis in Figs. \ref{figure4} - \ref{figure9} 
(where the normalized variable is now $\sim \lambda_{D, d}/\lambda \equiv k/k_{D, d}$).
\end{thereferences}
\end{document}